\documentclass[pra,twocolumn,superscriptaddress,showpacs,amsmath,amstex,amssymb,citeautoscript]{revtex4-1}

\usepackage[T1]{fontenc}
\usepackage[utf8]{inputenc}
\usepackage{lipsum}
\usepackage{amsmath}
\usepackage{amssymb}
\usepackage{bbm}
\usepackage{braket}
\usepackage{xcolor}
\usepackage{pifont}
\usepackage[mathscr]{euscript}
\usepackage[shortlabels]{enumitem}
\usepackage{graphicx}
\usepackage{lipsum}
\allowdisplaybreaks
\usepackage{float}
\usepackage{graphicx}
\usepackage{dsfont}
\usepackage{comment}
\usepackage[percent]{overpic}
\usepackage[colorlinks=true]{hyperref}  
\hypersetup{
    bookmarks=true,         
    unicode=false,          
    pdftoolbar=true,        
    pdfmenubar=true,        
    pdffitwindow=false,     
    pdfstartview={FitH},    
    pdftitle={Exactly Solvable Models Hosting Altermagnetic Quantum Spin Liquids},    
    pdfauthor={Sobral et al.},     
    pdfsubject={},   
    pdfcreator={},   
    pdfproducer={}, 
    pdfkeywords={} {} {}, 
    pdfnewwindow=true,      
    colorlinks=true,       
    linkcolor=blue, 
    citecolor=blue,        
    filecolor=magenta,      
    urlcolor=blue           
}

\newcommand{\equref}[1]{Eq.~(\ref{#1})}
\newcommand{\equsref}[2]{Eqs.~(\ref{#1}) and (\ref{#2})}
\newcommand{\secref}[1]{Sec.~\ref{#1}}
\newcommand{\figref}[1]{Fig.~\ref{#1}}
\newcommand{\refcite}[1]{Ref.~\onlinecite{#1}}

\newcommand{\appref}[1]{Appendix~\ref{#1}}

\newcommand{\pdagger}{{\phantom{\dagger}}}

\renewcommand{\approx}{\simeq}

\renewcommand{\vec}[1]{\boldsymbol{#1}}

\definecolor{wrongultramarine}{rgb}{1,0.5,0}

\usepackage{tikz}
\usepackage{amssymb}

\newcommand{\squarecross}[1][0]{%
  \tikz[baseline=-0.6ex, scale=0.1]{
    \begin{scope}[rotate=#1]
      \draw[line width=0.4pt] (-1,-1) rectangle (1,1);
      \draw[line width=0.4pt] (-1,-1) -- (1,1);
      \draw[line width=0.4pt] (-1,1) -- (1,-1);
    \end{scope}
  }%
}

\newcommand{\squarecrosscoupl}[1][0]{%
  \tikz[baseline=-0.6ex, scale=0.1]{
    \begin{scope}[rotate=#1]
      \draw[line width=0.4pt] (-1,-1) rectangle (1,1);
      \draw[line width=1pt] (-1,-1) -- (1,1);
      \draw[line width=1pt] (-1,1) -- (1,-1);
    \end{scope}
  }%
}

\newcommand{\emptysquare}[1][0]{%
  \tikz[baseline=-0.6ex, scale=0.1]{
    \begin{scope}[rotate=#1]
      \draw[line width=0.4pt] (-1,-1) rectangle (1,1);
    \end{scope}
  }%
}
\newcommand{\emptysquaresmall}[1][0]{%
  \tikz[baseline=-0.6ex, scale=0.07]{
    \begin{scope}[rotate=#1]
      \draw[line width=0.4pt] (-1,-1) rectangle (1,1);
    \end{scope}
  }%
}
\newcommand{\squaretriangle}[2][0]{%
  \tikz[baseline=-0.6ex, scale=0.1]{
    \begin{scope}[rotate=#1]
      \draw[line width=0.4pt] (-1,-1) rectangle (1,1);
      \draw[line width=0.4pt] (-1,-1) -- (1,1);
      \draw[line width=0.4pt] (-1,1) -- (1,-1);
      \ifnum#2=1 \fill (-1,1) -- (1,1) -- (1,-1) -- cycle; \fi 
      \ifnum#2=2 \fill (-1,-1) -- (1,-1) -- (1,1) -- cycle; \fi 
      \ifnum#2=3 \fill (-1,-1) -- (-1,1) -- (1,1) -- cycle; \fi 
      \ifnum#2=4 \fill (-1,-1) -- (-1,1) -- (1,-1) -- cycle; \fi 
    \end{scope}
  }%
}

\newcommand{\emptyoctagon}[1][0]{%
  \tikz[baseline=-0.6ex, scale=0.1]{
    \begin{scope}[rotate=#1]
      \draw[line width=0.4pt] 
        (1,0.414) -- (0.414,1) -- (-0.414,1) -- (-1,0.414) -- 
        (-1,-0.414) -- (-0.414,-1) -- (0.414,-1) -- (1,-0.414) -- cycle;
    \end{scope}
  }%
}

\newcommand{\halfoctagon}[1][0]{%
  \tikz[baseline=-0.6ex, scale=0.1]{
    \begin{scope}[rotate=#1]
      \coordinate (A) at (1,0.414);
      \coordinate (B) at (0.414,1);
      \coordinate (C) at (-0.414,1);
      \coordinate (D) at (-1,0.414);
      \coordinate (E) at (-1,-0.414);
      \coordinate (F) at (-0.414,-1);
      \coordinate (G) at (0.414,-1);
      \coordinate (H) at (1,-0.414);

      \fill[black!80] (A) -- (B) -- (C) -- (D) -- (0,0) -- (H) -- cycle;

      \draw[line width=0.4pt] (A) -- (B) -- (C) -- (D) -- (E) -- (F) -- (G) -- (H) -- cycle;
    \end{scope}
  }%
}

\begin{document}
\title{Exactly Solvable Models Hosting Altermagnetic Quantum Spin Liquids}
\author{João Augusto Sobral}
\affiliation{Institute for Theoretical Physics III, University of Stuttgart, 70550 Stuttgart, Germany}
\author{Pietro M. Bonetti}
\affiliation{Department of Physics, Harvard University, Cambridge MA 02138, USA}
\author{Subrata Mandal}
\affiliation{Institute for Theoretical Physics III, University of Stuttgart, 70550 Stuttgart, Germany}
\author{Mathias S.~Scheurer}
\affiliation{Institute for Theoretical Physics III, University of Stuttgart, 70550 Stuttgart, Germany}

\begin{abstract}
We construct spin-3/2 and spin-7/2 models on the square-octagon and checkerboard lattices that are exactly solvable with Majorana representations. They give rise to spin-liquid phases with full spin-rotation and lattice-translational symmetries but broken time-reversal symmetry. Although non-zero on elementary plaquettes, the net orbital magnetic moment is guaranteed to vanish as a result of point symmetries; due to the analogy to long-range ordered altermagnets, these types of phases were dubbed altermagnetic spin liquids in [\href{https://journals.aps.org/prresearch/abstract/10.1103/PhysRevResearch.7.023152}{Phys.~Rev.~Research \textbf{7}, 023152}]. For the spin-3/2 model, we find that a $g$-wave altermagnetic spin liquid emerges as the unique ground state. In contrast, the spin-7/2 model exhibits a significantly richer phase diagram, involving different types of chiral spin liquids competing with a $d$-wave altermagnetic spin liquid. Finally, we identify and characterize the topological and non-topological excitations, illustrating the rich physics of altermagnetic spin liquids resulting from the interplay of non-trivial topological and symmetry aspects of this novel phase of matter.
\end{abstract}

\maketitle

\section{Introduction}
Altermagnetism refers to a class of magnetic states with vanishing net magnetization (unlike ferromagnets) which is enforced not as a result of translational symmetry combined with time-reversal (like in anti-ferromagnets) but follows from a combination of the latter with a point-group operation \cite{smejkal_emerging_2022}. Although initially driven by their relevance to spintronics, altermagnets have also become subject of intense research in the field of strongly correlated physics and superconductivity \cite{Jungwirth2025Aug, Liu2025Oct}.
For instance, the fate of these long-range ordered altermagnetic phases in regimes dominated by strong quantum fluctuations---particularly in proximity to, or within, quantum spin liquids (QSLs) \cite{Savary_2017, Zhou2017Apr}---has been recently explored in distinct scenarios \cite{Sobral2025May,Vijayvargia2025Oct, Neehus2025Apr}.

\refcite{Sobral2025May} showed that a variety of exotic QSLs can descend from long-range ordered parent altermagnets, with some retaining the same (alter-)magnetic point-group symmetries in experimentally accessible observables, in both metallic and insulating regimes. Notably, the electronic spectral functions of the doped system can still exhibit the characteristic altermagnetic momentum-dependent splitting, even though the topologically ordered phase maintains full SU(2) spin-rotational invariance. Among the identified phases, the “orbital altermagnet” (see also \cite{yu2024altermagnetism,PhysRevLett.134.146001,2025arXiv250926596C,2025arXiv251000509P} for a discussion of orbital altermagnetism in different contexts) exhibits a pattern of orbital currents arranged in an altermagnetic fashion. The descendant QSL inherits a corresponding structure in its emergent gauge-flux configuration and scalar spin chiralities, leading to its designation as an orbital altermagnetic quantum spin liquid (OAMSL).

Starting from a Kitaev-type \cite{KITAEV20062} gapless $\mathbb{Z}_{2}$ QSL defined on a bilayer square–octagon lattice, \refcite{Vijayvargia2025Oct} showed that increasing the strength of additional interactions that commute with both intra- and interlayer flux operators of the original model drives a first-order transition into a $\vec{q}=0$ ordered state. In this phase, a finite local symmetry-breaking order parameter coexists with remnant global $\mathbb{Z}_{2}$ topological order inherited from the parent gapless QSL—an example of magnetic fragmentation \cite{Vijayvargia2023Jun, Brooks-Bartlett2014Jan, Petit2016Aug}. This coexistence motivated the designation of the phase as a $\mathbb{Z}_{2}$ topological altermagnet. The bands of the fractionalized excitations exhibit a $d$-wave momentum-dependent spin-splitting protected by the combination of a diagonal mirror symmetry and time reversal. 

A related phenomenon was identified in an exactly solvable $\mathbb{Z}_{2}$ quantum-spin(-orbital) liquid on the square lattice \cite{Neehus2025Apr}. Small perturbations (compared to the flux gap) which obey altermagnetic projective symmetry requirements, and that do not break the integrability of the model, lead to an effective low-energy theory for fermionic partons whose particle–hole asymmetry generates momentum-dependent band splittings that are analogous to altermagnets. These were dubbed fractionalized altermagnets, akin to fractionalized antiferromagnets \cite{Nayak2000Aug,Senthil2004Jan, Ghaemi2006Feb}, and were classified according to their projective symmetry group \cite{Wen2002Apr}.

Taken together, these advances reveal that altermagnetic symmetry breaking can appear in QSLs either as a descendant of an ordered parent phase with fully preserved spin-rotation symmetry and potentially additional altermagnetic scalar spin chiralities \cite{Sobral2025May} or through coexistence of an altermagnetic spin texture with remnant $\mathbb{Z}_2$ topological order \cite{Vijayvargia2025Oct,Neehus2025Apr}.

Motivated by these developments, and building on the Kitaev \cite{KITAEV20062} and Yao–Kivelson \cite{Yao2009May} constructions, we introduce two exactly solvable models that intrinsically host OAMSLs in the sense defined in \refcite{Sobral2025May}. The ground state of the spin-3/2 model is a $g$-wave OAMSL, in sharp contrast with the more complex phase diagram arising for the spin-7/2 model. We further address the different types of excitations in the flux sector.

\section{Spin-3/2 model}
As already outlined above, our goal is to construct an exactly solvable model hosting an OAMSL, which is characterized by the following properties: while spin-rotation invariance is preserved, time-reversal symmetry is spontaneously broken. However, as opposed to a chiral spin liquid (which can be thought of as the ferromagnetic analog of the OAMSL), there are still residual magnetic point group symmetries that guarantee that the associated net (orbital) magnetic moment vanishes. Similarly to the difference between an antiferromagnet and an altermagnet, we emphasize that the vanishing of the moment is \textit{not} the consequence of the product of time-reversal and translation; the OAMSLs we will discuss here all preserve the translational symmetry of the underlying lattice. 

\begin{figure}[b]
    \centering
    \raisebox{-10pt}{%
        \begin{overpic}[width=.575\linewidth]{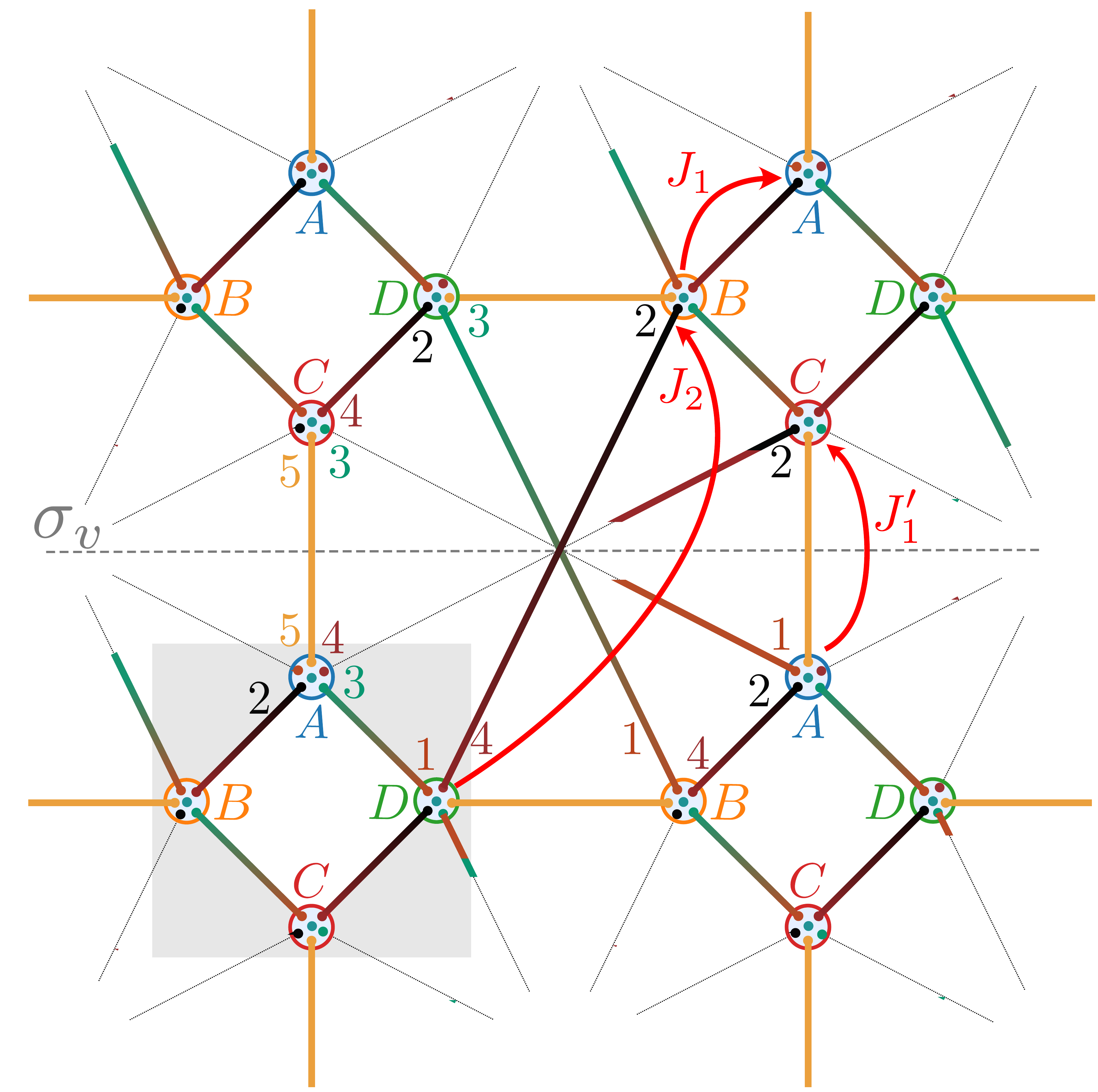}
            \put(0,84){(a)}
        \end{overpic}
    }\hspace{2mm}%
    \begin{overpic}[width=.35\linewidth]{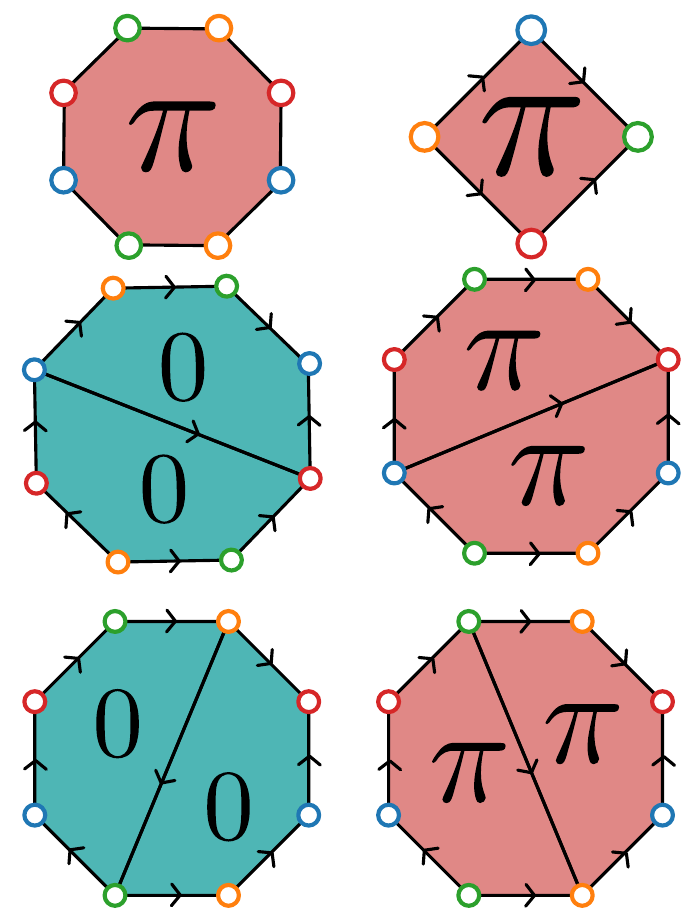}
        \put(-10,95){(b)}
    \end{overpic}

    \caption{Schematic representation of the spin-3/2 model defined in \equsref{eq:spin3/2model}{eq:HF3/2}. (a) The chosen unit cell is illustrated with a gray square. The horizontal gray line indicates the $\sigma_v$ reflection plane. The blue, orange, red, and green circles denote lattice sites belonging to the $A$, $B$, $C$ and $D$ sublattices, respectively. The red arrows denote a particular instance of the interactions with coefficients $J_1$, $J_2$ and $J_5$. The numbers around each lattice site indicate which $\Gamma$-matrix appears for the given sublattice site in the interaction term along the bond on which the number is located. 
    (b) Gauge flux structure of the $g$-wave OAMSL. The elementary plaquettes include one square, one octagon, and pentagons inside the latter, with fluxes $0$, $\pi$  and  $\pi$ respectively. The arrows indicate the direction in which the respective $\hat{u}_{i_1,i_2}^{\mu_1,\mu_2}$ is positive.}
    \label{fig:firstmodel}
\end{figure}

In the spirit of the Kitaev \cite{KITAEV20062} and Yao–Kivelson \cite{Yao2009May} constructions, the coordination number of the lattice determines the ``length'' $S$ of the spin operators on every site. Furthermore, to break time-reversal symmetry, we require elementary loops with an odd number of sites \cite{PhysRevLett.99.247203,Chua2011May}. Focusing on tetragonal two-dimensional systems, as in Ref. \cite{Sobral2025May}, the checkerboard lattice seems like a natural starting point to ensure the second constraint. However, the coordination number is six, which requires $S=7/2$. While we will discuss such a model in \secref{Sec:Spin72Model} below, we will first reduce the coordination number to five by working on the square-octagon lattice (see \figref{fig:firstmodel}a). In this case, $S=3/2$ suffices as its four-dimensional onsite Hilbert space allows for five $\Gamma^a$ matrices which obey the Clifford algebra $\{\Gamma^a,\Gamma^b\}=2\delta^{ab}\mathbb{I}_{4}$. The Hamiltonian is defined by
\begin{widetext}
\begin{equation}
\label{eq:spin3/2model}
\begin{gathered}
{\cal H}_{3/2} = \,\sum_{\vec{R}}   \left[J_1 \left(\Gamma^3_{\vec{R},A} \Gamma^1_{\vec{R},D} + \Gamma^4_{\vec{R},B} \Gamma^2_{\vec{R},A} 
+ \Gamma^2_{\vec{R},D} \Gamma^4_{\vec{R},C} + \Gamma^1_{\vec{R},C} \Gamma^3_{\vec{R},B}   \right) + J_1^{\prime} \left( \Gamma^5_{\vec{R},D} \Gamma^5_{\vec{R}_x,B} + \Gamma^5_{\vec{R}_y,C} \Gamma^5_{\vec{R},A} \right)\right]+\\
+ \sum_{\vec{R}}J_2 \left(  \Gamma^3_{\vec{R}_y,C} \Gamma^1_{\vec{R}_x,A} + \Gamma^4_{\vec{R},A} \Gamma^2_{\vec{R}_{x+y},C}  
+ \Gamma^4_{\vec{R},D} \Gamma^2_{\vec{R}_{x+y},B} + \Gamma^1_{\vec{R}_x,B} \Gamma^3_{\vec{R}_y,D} \right),
\end{gathered} 
\end{equation}
\end{widetext}
with nearest-neighbor couplings $J_{1}$ along the edges of the square $\emptysquare[45]$, $J_{1}^{\prime}$ along the horizontal and vertical edges of the octagon $\emptyoctagon[0]$, and second-neighbor interactions $J_{2}$ inside the octagons. Here, $\vec{R}$ labels the Bravais lattice sites (gray square in \figref{fig:firstmodel}a) and the shorthands $\vec{R}_{\delta}:= \vec{R}+\vec{e}_\delta$, $\delta = x, y$, and $\vec{R}_{x+y}:= \vec{R}+\vec{e}_x+\vec{e}_y$ are used to refer to neighboring unit cells. The second subscript of the $\Gamma$-matrices refers to the four sublattices $\alpha=A,B,C,D$ in each unit cell.

There are many different ways of representing the $\Gamma^a$  matrices acting on the four-dimensional Hilbert space on each site (suppressing the site index here). For instance, they can be expressed in terms of two spin-$1/2$ or two qubits per site. Denoting the Pauli matrices by $\sigma^j$, one possible representation is given by 
\begin{equation}
\label{eq:spin-3/2map}
\begin{gathered}
 \Gamma^1=\sigma^z \otimes \sigma^y, \, \,\Gamma^2=\sigma^z \otimes \sigma^x, \, \Gamma^3=\sigma^y \otimes \mathbb{I}_2, \\
 \Gamma^4=\sigma^x \otimes \mathbb{I}_2, \, \Gamma^5=\sigma^z \otimes \sigma^z.
\end{gathered}
\end{equation}
Alternatively, the $\Gamma^a$ matrices can also be expressed as spin-3/2 operators $S^{j}$ (see \appref{app:higherorder}), such that  \equref{eq:spin3/2model} can be reinterpreted as a Hamiltonian with interacting quadrupole moments \cite{Demler1999Jan, Murakami}.

The point group of the lattice is $C_{4v}$, which is generated by 90$^\circ$ anti-clockwise rotations ($C_{4z}$) and reflections about the $x$-axis ($\sigma_v$), as indicated in \figref{fig:firstmodel}a.  From \equref{eq:spin3/2model}, the symmetry operations act on the $\Gamma^a$ matrices as follows 

\begin{equation}
\label{eq:sym32}
\begin{gathered}
C_{4 z}: (\Gamma^1, \Gamma^2, \Gamma^3, \Gamma^4) \mapsto (\Gamma^2, \Gamma^3, \Gamma^4, \Gamma^1), 
\, \Gamma^5 \mapsto -\Gamma^5,  
 \\
\sigma_v: \Gamma^1 \leftrightarrow \Gamma^2, \, 
\Gamma^3 \leftrightarrow \Gamma^4,\, 
\Gamma^5 \leftrightarrow \Gamma^5.
\end{gathered}
\end{equation}
Noting that the Hamiltonian in Eq.~\eqref{eq:spin3/2model} remains invariant under both symmetry operations, and we can conclude that its point group is $C_{4v}$. Furthermore, time-reversal is represented by the anti-unitary operator $\Theta$ with action $\Theta \Gamma^j \Theta^\dagger = \Gamma^j$. In the representation stated in \equref{eq:spin-3/2map}, it is given by $\Theta = \sigma^{x} \otimes \sigma^y \mathcal{K}$, where $\mathcal{K}$ is the complex conjugation operator.
It is clearly also a symmetry of $\mathcal{H}_{3/2}$ in \equref{eq:spin3/2model}.

To diagonalize the Hamiltonian~\eqref{eq:spin3/2model}, we employ a standard Majorana decomposition of the $\Gamma^{a}$ matrices, introducing six Majorana fermions per site (shown as colored dots in \figref{fig:firstmodel}a). At each site $i=(\vec{R},\alpha)$, we define operators $c_i^\mu$ and $d_i$ satisfying
$\Gamma_i^\mu = i\, c_i^\mu d_i \; (\mu = 1,\dots,5).$

As in the Kitaev model \cite{KITAEV20062}, the $c_i^\mu$ Majoranas are immobile and generate a static $\mathbb{Z}_2$ gauge-field background through the bond variables $\hat u_{i_1,i_2}^{\mu_1,\mu_2} = i c_{i_1}^{\mu_1} c_{i_2}^{\mu_2}$. 
The mobile $d_i$ Majoranas then form the dispersive bands. Each Majorana fermion contributes a nominal dimension of $\sqrt{2}$ such that the six Majoranas span an enlarged eight-dimensional Fock space, in contrast to the four-dimensional Hilbert space of a spin-$3/2$. Since $\Gamma^1_i\Gamma^2_i\Gamma^3_i\Gamma^4_i\Gamma^5_i=-\mathbb{I}_4 \;, \forall i$, all allowed physical states $\ket{\Psi}$ must satisfy a local parity constraint given by 
\begin{eqnarray}
D^{(3/2)}_i|\Psi\rangle\equiv\left[-ic^1_ic^2_ic^3_ic^4_ic^5_id_i\right] |\Psi\rangle =|\Psi\rangle.
\label{eq:physicalconstraint}
\end{eqnarray}
In practice, this needs to be enforced by the projection operator 
$
|\Psi\rangle= P^{(3/2)}|\psi\rangle \equiv \prod_{i}\Big[(1+D^{(3/2)}_i)/2\Big]|\psi\rangle,$
which ensures $D^{(3/2)}_i|\Psi\rangle=|\Psi\rangle$ for all sites. Explicitly,
$
P^{(3/2)}=\Big[1+\sum_i D^{(3/2)}_i+\sum_{i_1<i_2}D^{(3/2)}_{i_1}D^{(3/2)}_{i_2}+\cdots+\prod_i D^{(3/2)}_i\Big]/2^N, 
$
where $N$ is the total number of sites \cite{Yao2009May}. The constraint \eqref{eq:physicalconstraint} can be naturally rewritten in terms of the $\mathbb{Z}_{2}$ link variables $ u_{i_1,i_2}^{\mu_1,\mu_2}$ and ultimately depends on the lattice boundary conditions (see \appref{app:physicalstates} for further details) \cite{Pedrocchi2011Oct,Zschocke2015Jul}.

After rewriting in terms of the Majorana fermions as outlined above, the Hamiltonian then takes the form 
\begin{widetext} \begin{equation} \label{eq:HF3/2} \begin{aligned} \mathcal{H}_{3 / 2}=&i \sum_{\vec{R}} \left(J_1 \sum_{(\alpha, \beta, \mu, \nu) \in \mathcal{P}_{\text {intra }}} \hat{u}_{(\vec{R}, \alpha),(\vec{R}, \beta)}^{\mu \nu} d_{\vec{R}, \alpha}^{\pdagger} d_{\vec{R}, \beta}^{\pdagger} + J_1^{\prime} \sum_{(\delta_1,\delta_2,\alpha, \beta) \in \mathcal{P}^{\prime}} \hat{u}_{(\vec{R}_{\delta_1}, \alpha),(\vec{R}_{\delta_2}, \beta)}^{55} d_{\vec{R}_{\delta_1}, \alpha}^{\pdagger} d_{\vec{R}_{\delta_2}, \beta}^{\pdagger} \right)\\ & \quad+\quad i \sum_{\vec{R}} J_2 \sum_{(\delta_1, \delta_2, \alpha, \beta, \mu, \nu) \in \mathcal{P}_{\text{inter}}} \hat{u}_{(\vec{R}_{\delta_1}, \alpha),(\vec{R}_{\delta_2}, \beta)}^{\mu \nu} d_{\vec{R}_{\delta_1}, \alpha}^{\pdagger} d_{\vec{R}_{\delta_2}, \beta}^{\pdagger} + \text {h.c.}, \end{aligned} \end{equation} \end{widetext}
with $\mathcal{P}_{\text{intra}}$, $\mathcal{P}^{\prime}$, and
$\mathcal{P}_{\text{inter}}$ denoting the sets of intra-unit-cell,
primed, and inter-unit-cell bonds, respectively, defined as (see \figref{fig:firstmodel}a)
\begin{align}
\mathcal{P}_{\text{intra}} &=
\{(A,D,3,1), (B,A,4,2), \notag (C,B,2,4), (D,C,1,3)\}, \\
\mathcal{P}^{\prime} &=
\{(0,x,D,B), (y,0,C,A)\},  \notag \\
\mathcal{P}_{\text{inter}} &=
\{(y,x,C,A,3,1), (0,x+y,A,C,4,2), \\
&\quad\;\;\;(0,x+y,D,B,4,2), (x,y,B,D,1,3)\}.\notag 
\end{align}
From their definition, the Hermitian bond operators $\hat u_{i_1,i_2}^{\mu_1,\mu_2}$ satisfy
$\left( \hat u_{i_1,i_2}^{\mu_1,\mu_2} \right)^2 = 1$ and commute with the Hamiltonian,
so they are conserved quantities with eigenvalues $\pm 1$. Thus, Eq.~\eqref{eq:HF3/2} describes free Majorana fermions $d_i$ moving in a static $\mathbb{Z}_2$ gauge-field background specified by the eigenvalues of $\{\hat u_{i_1,i_2}\}$. 
Since the $\{\hat u_{i_1,i_2}\}$ are not gauge invariant, we instead focus on closed loops of them, which define conserved and gauge invariant fluxes through the areas they enclose. As they serve as building blocks of fluxes through larger areas, we here focus on the fluxes through the elementary plaquettes (squares, octagons and pentagons inside the latter) via the Wilson loop operators $\hat W_{\text{EP}} = \prod_{\langle ij\rangle \in \text{EP}} \hat u_{ij}$, where $\text{EP} \in \{ \emptyoctagon[0],  \emptysquaresmall[45],  \halfoctagon[45], \halfoctagon[-135], \halfoctagon[0], \halfoctagon[-180], \halfoctagon[-270],\halfoctagon[-90], \halfoctagon[315],\halfoctagon[135]\}$. Throughout this work, we will follow the conventions that those loops are taken in a clockwise orientation. 

For finite couplings $|J_{\{1,2,3\}}|\le 10$, we find numerically that the ground state has $\pi$-fluxes on the square and octagon  plaquettes, with a staggered orbital flux pattern on the eight pentagons, as summarized graphically in \figref{fig:firstmodel}b. This can be contrasted with the square-octagon monolayer limit in \refcite{Vijayvargia2025Oct}, where a ground state with uniform $\pi$ fluxes on the squares and octagons is found, which is a consequence of Lieb's theorem \cite{Lieb1989Mar}. Importantly, they consider a distinct model from \equref{eq:spin3/2model} with no analogue of our $J_{2}$ couplings, which is essential to bipartiteness.

Importantly, as a result of the anti-unitary nature of time-reversal, it must hold $\Theta c_i^\mu \Theta^\dagger = c_i^\mu$ and $\Theta d_i \Theta^\dagger = -d_i$ (modulo any additional $\mathbb{Z}_2$ gauge transformation, irrelevant for our discussion here). This further implies that $\Theta \hat u_{i_1,i_2}^{\mu_1,\mu_2} \Theta^\dagger = -\hat{u}_{i_1,i_2}^{\mu_1,\mu_2}$ and, hence, all Wilson loop operators with an even and odd number of edges are even and odd under time-reversal, respectively. Consequently, the flux configuration in \figref{fig:firstmodel}b with $0$ and $\pi$ swapped in the eight pentagons is degenerate with the one shown. Favoring one over the other in the ground state will spontaneously break time-reversal symmetry. Further inspection of the flux pattern reveals that it preserves $\sigma_v$ and $C_{4z}$; these symmetries ensure that the net magnetization vanishes and the associated time-reversal-odd altermagnetic order parameter transforms under $A_1$ of $C_{4v}$. Since the lowest-order contribution in a multipole expansion of the magnetization is of the form $xy(x^2-y^2)$, one can think of it as a $g$-wave altermagnet \cite{smejkal_emerging_2022, banerjeeAltermagneticSuperconductingDiode2024, Fernandes2024Jan}. Recalling that spin rotation invariance is preserved in the QSL ground state and this form of time-reversal symmetry breaking is in the orbital channel, we refer to it as a $g$-wave OAMSL.

\begin{figure*}[htp!]
    \centering
    \includegraphics[width=1\linewidth]{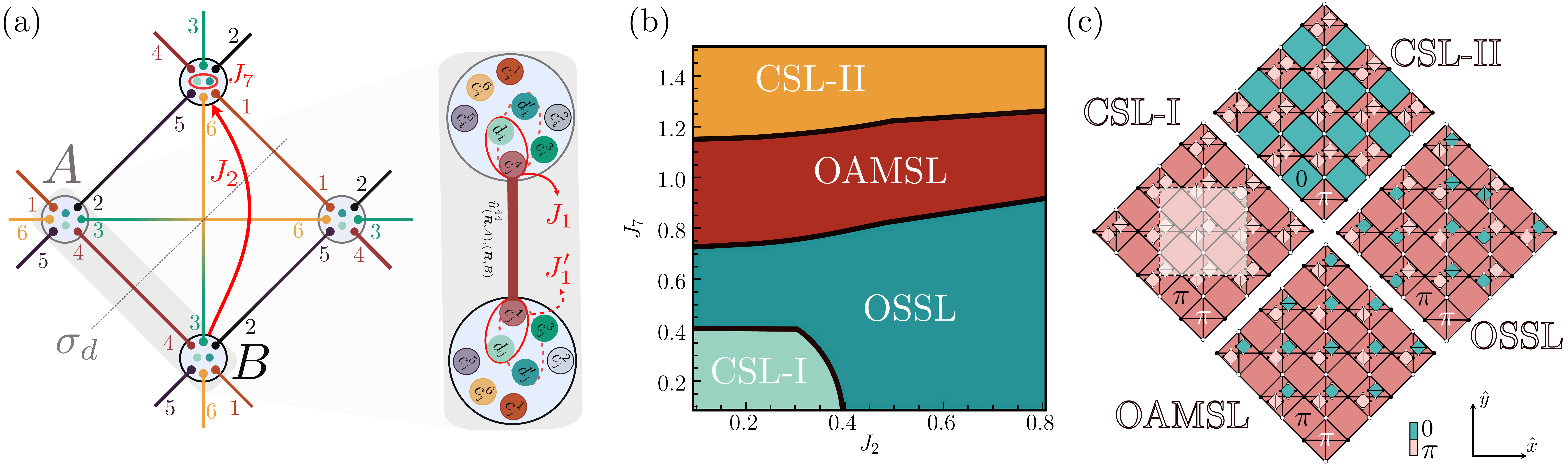}
    \caption{Schematic representation of the spin-7/2 model in \equsref{eq:spin7/2model}{eq:spin7/2modelmajor}. 
(a) Lattice structure with the $\sigma_{d}$ reflection symmetry indicated by the diagonal dashed gray line. 
As in \figref{fig:firstmodel}a, the numbers around each lattice site indicate which $\gamma^{a}$-matrix appears in the interaction term for the given sublattice site along the corresponding bond. 
Each colored dot represents one Majorana fermion, as defined in \equref{eq:majorana7/2} and also shown in the zoomed-in unit cell. 
The red lines and symbols illustrate representative interactions with coefficients $J_{1}$, $J_{2}$, and $J_{7}$. 
(b) Phase diagram as a function of $J_{2}$ and $J_{7}$ for  $J_{2}^{\prime}=J_{2}$ and $J_{1}^{\prime}=J_{1}=1$. In this regime, there are two chiral spin liquids (CSL), an orbital altermagnetic spin liquid (OAMSL) and an orbital stripy spin liquid (see main text for further details about each phase). 
(c)~Flux configurations for each phase shown in (b). 
Elementary plaquettes (empty and crossed squares, and triangles inside the latter) are colored according to their gauge flux (oriented clockwise), with values $0$ (green) or $\pi$ (red).
The white square in the CSL-I phase indicates the unit cell size considered in our analysis. }
    \label{fig:secondmodel}
\end{figure*}

\section{Spin-7/2 model}\label{Sec:Spin72Model}
As already anticipated above, we next continue with a model on the checkerboard lattice (see \figref{fig:secondmodel}a). As each site is coupled to six neighbors, we need at least six anticommuting matrices. This can be realized with an eight-dimensional irreducible representation of the Clifford algebra, which leads to a total of seven $8 \times 8$ matrices $\gamma^a$ ($a=1,2\dots, 7$) obeying $\{\gamma^a,\gamma^b\}=2\delta^{ab}\mathbb{I}_{8}$. In particular, one possible three-qubit per site (or three spin-$1/2$) representation is given by
\begin{equation}
\label{eq:spin-7/2map}
\begin{aligned}
 \gamma^1&=\sigma^x \otimes \sigma^{z} \otimes \sigma^{x},\,  \\\gamma^2&=\sigma^x \otimes \sigma^{y} \otimes \mathbb{I}_2, \, \\ \gamma^3&=\sigma^x \otimes \sigma^{x} \otimes \mathbb{I}_2, 
  \\ \gamma^4&=\sigma^z \otimes \mathbb{I}_2 \otimes \mathbb{I}_2,\,  \\\gamma^5&=\sigma^y \otimes \mathbb{I}_2 \otimes \mathbb{I}_2, \, \\\gamma^6&=\sigma^x \otimes \sigma^{z} \otimes \sigma^{y},\,\\ \gamma^7&=\sigma^x \otimes \sigma^{z} \otimes \sigma^{z}.
\end{aligned}
\end{equation}
Instead of just thinking of the local on-site degrees of freedom as some eight-level systems, one can also construct them as polynomials of spin-7/2 operators. We refer to \appref{app:higherorder} for details.
With one $\gamma$ matrix more than the number of neighbors, we can also include additional exchange couplings involving the commutators $\gamma^{ab}=\left[\gamma^{a},\gamma^{b}\right]/2i$, without spoiling the exact solvability of the model in terms of Majorana fermions. The explicit Hamiltonian we study is defined as 
\begin{widetext}
\begin{multline}
\mathcal{H}_{7/2}=\, \sum_{\vec{R}}
\Bigl[J_1\bigl(\gamma_{\vec{R}, A}^4 \gamma_{\vec{R}, B}^4
+\gamma_{\vec{R}, A}^2 \gamma_{\vec{R}_y, B}^5
+\gamma_{\vec{R}_x, A}^1 \gamma_{\vec{R}_y, B}^1
+\gamma_{\vec{R}, B}^2 \gamma_{\vec{R}_x, A}^5\bigr) 
+J_2\bigl(\gamma_{\vec{R}, A}^3 \gamma_{\vec{R}_x, A}^6
+\gamma_{\vec{R}, B}^3 \gamma_{\vec{R}_y, B}^6\bigr)- J_7\bigl(\gamma_{\vec{R}, A}^7+\gamma_{\vec{R}, B}^7\bigr)\Bigr] +\\
\quad +\; \sum_{\vec R}\Bigl[J_1^{\prime}\bigl(\gamma_{\vec{R}, A}^{47}\gamma_{\vec{R}, B}^{47}
+\gamma_{\vec{R}, B}^{27} \gamma_{\vec{R}_y, B}^{57}
+\gamma_{\vec{R}_x, A}^{17} \gamma_{\vec{R}_y, B}^{17}
+\gamma_{\vec{R}, B}^{27} \gamma_{\vec{R}_x, A}^{57}\bigr) +J_2^{\prime}\bigl(\gamma_{\vec{R}, B}^{37}\gamma_{\vec{R}_y, B}^{67}
+\gamma_{\vec{R}, A}^{37} \gamma_{\vec{R}_x, A}^{67}\bigr)
\Bigr],
\label{eq:spin7/2model}\end{multline}
\end{widetext}
with first-neighbor interactions $(J_{1}, J_1')$ on the diagonal sides of the squares \emptysquare[45], diagonal $(J_{2},J_2^{\prime})$ couplings on the crossed squares \squarecrosscoupl[45], and on-site $J_{7}$ couplings. We use the same notation as above with $\vec{R}$ labeling the Bravais lattice sites, the shorthand $\vec{R}_{x,y}:= \vec{R}+\vec{e}_{x,y}$, and the second subscript referring to the two sublattices $\alpha=A,B$ in each unit cell. As we will see below, the primed exchange couplings $J_{1,2}^{\prime}$  are necessary to make the model sufficiently generic and to remove an extensive degeneracy, and $J_7$ will be instrumental in stabilizing non-trivial orbital textures over triangle fluxes.  

If we demand that the representations of the transformations in the $8 \times 8$ space of the $\gamma^a$ are site independent, this model will havefewer symmetries than the previous one. In that case, the point group is now $C_{2v}$, which is generated by 180$^\circ$ anti-clockwise rotations ($C_{2z}$) and reflections about the diagonal axis ($\sigma_d$) indicated in \figref{fig:secondmodel}a. The symmetry operations act on the $\gamma^a$ matrices as \begin{align}
\label{eq:sym72}
C_{2 z}:\;& \gamma^1 \mapsto -\gamma^4, \, \gamma^4 \mapsto \gamma^1,\,  (\gamma^2, \gamma^3) \leftrightarrow (\gamma^5, \gamma^6), 
\;  \gamma^7 \mapsto \gamma^7, \notag \\ 
 \\
\sigma_d:\;& \gamma^j \rightarrow \gamma^j \; (j=1,\dots 7).\notag  \; 
\end{align}
Although the lattice also has a $\sigma_v$ symmetry and a 90$^\circ$ rotational symmetry $C_{4z}$, these are not compatible with placing the $\gamma^a$ on the bonds and site-independent representations: the exact solvability requires that the indices $a$ of the $\gamma^a$ matrices ``meeting'' at any vertex must all be different. Applying this to the vertex $B$ in \figref{fig:secondmodel}a, we conclude that three disjoint pairs of the seven $\gamma$ matrices must swap under $\sigma_v$ (for instance, $\gamma^1 \leftrightarrow \gamma^2$, $\gamma^3 \leftrightarrow \gamma^4$, $\gamma^5 \leftrightarrow \gamma^6$). However, applying it to the vertex $A$ also requires that at least two $\gamma$ matrices must map to themselves under $\sigma_v$, leading to a contradiction. Since $\sigma_v$ would follow from $\sigma_d$ and $C_{4z}$, it cannot be possible to realize $C_{4z}$ either. We note, however, that there is a site-dependent representation, i.e., $\gamma^a_i \rightarrow U_g(i) \gamma^a_{gi} U^\dagger_g(i)$ of the symmetries $g=C_{4z}, \sigma_v$ leaving the Hamiltonian~(\ref{eq:spin7/2model}) invariant (see \appref{app:higherorder}). Although not essential for the definition of the OAMSL, we will use this below to further characterize the flux patterns of the different spin liquids.

To diagonalize the Hamiltonian \eqref{eq:spin7/2model}, we employ another Majorana decomposition given by 
\begin{equation}
\label{eq:majorana7/2}
    \gamma_{i}^{a}=ic_{i}^{a}d_{i},\; \gamma_{i}^{a 7}=ic_{i}^{a}d_{i}^{\prime}\; \text{and} \; \gamma_{i}^{7}=id_{i}^{\prime}d_{i}^{\phantom{'}},
\end{equation}
where $a=1,\dots,6$ and we again used the multi-index notation $i=(\vec{R},\alpha)$. These eight Majorana fermions are indicated as colored dots on each site in \figref{fig:secondmodel}a. Although the $c_i^\mu$ fermions are still immobile, there are now two mobile $d_i$ and $d_i^{\prime}$ fermions. The enlarged sixteen-dimensional Fock space---in contrast to the eight-dimensional Hilbert space of a spin-$7/2$---needs the constraint $D^{(7/2)}_i|\Psi\rangle\equiv\left[c^1_ic^2_ic^3_ic^4_ic^5_ic^6_id_i^{\pdagger}d_i^{\prime}\right] |\Psi\rangle =|\Psi\rangle$, arising from $i\gamma^1_i\gamma^2_i\gamma^3_i\gamma^4_i\gamma^5_i\gamma^6_i \gamma^7_i=\mathbb{I}_8$ (see \appref{app:physicalstates}). By introducing the new bond operators  $\hat u_{i_{1},i_{2}}^{\mu_1,\mu_2}=ic_{i_1}^{\mu_1}c_{i_2}^{\mu_{2}}$, the Hamiltonian in \equref{eq:spin7/2model} can be rewritten as
\begin{widetext} \begin{equation} \label{eq:spin7/2modelmajor} \begin{split} \mathcal{H}_{7 / 2}=&i \sum_{\vec{R}} \left[ \sum_{(\delta_1, \delta_2, \alpha, \beta, \mu, \nu) \in \mathcal{P}_{\text {\emptysquare[45]}}} \hat{u}_{(\vec{R}_{\delta_1}, \alpha),(\vec{R}_{\delta_2}, \beta)}^{\mu \nu} \left(J_{1}^{\pdagger}d_{\vec{R}_{\delta_1}, \alpha}^{\pdagger} d_{\vec{R}_{\delta_2}, \beta}^{\pdagger} +J_{1}^{\prime}d_{\vec{R}_{\delta_1}, \alpha}^{\prime} d_{\vec{R}_{\delta_2}, \beta}^{\prime}\right)-iJ_7\left(d_{\vec{R}, A}^{\prime}d_{\vec{R}, A}^{\pdagger}+d_{\vec{R}, B}^{\prime}d_{\vec{R}, B}^{\pdagger}\right)\right]+\\& 
+i \sum_{\vec{R}}  \sum_{(\delta_1, \delta_2, \alpha, \beta) \in \mathcal{P}_{\text {\squarecrosscoupl[45]}}} \hat{u}_{(\vec{R}_{\delta_1}, \alpha),(\vec{R}_{\delta_2}, \beta)}^{36} \left(J_{2}^{\pdagger}d_{\vec{R}_{\delta_1}, \alpha}^{\pdagger} d_{\vec{R}_{\delta_2}, \beta}^{\pdagger} +J_{2}^{\prime}d_{\vec{R}_{\delta_1}, \alpha}^{\prime} d_{\vec{R}_{\delta_2}, \beta}^{\prime}\right)+\text{h.c.}, \end{split} \end{equation} \end{widetext}
with $\mathcal{P}_{\emptysquare[45]}$ and $\mathcal{P}_{\squarecrosscoupl[45]}$ denoting the sets of first- and second-neighboring bonds, defined as (see \figref{fig:secondmodel}a)
\begin{align}
\mathcal{P}_{\emptysquare[45]}&=
\{(0, 0, A, B, 4, 4), (0, y, A, B, 2, 5),\notag \\& (x, y, A, B, 1, 1), (0, x, B, A, 2, 5)  \}, \notag\\
\mathcal{P}_{\squarecrosscoupl[45]} &=
\{(0,x, A,A), (0,y,B,B)\}.
\end{align}
This Hamiltonian  describes free $d_{i,\alpha }$ and $d_{i,\alpha}^{\prime}$ Majorana fermions. In this representation, the difference between the $J_{\{1,2\}}^{\pdagger}$ and $J_{\{1,2\}}^{\prime}$ couplings becomes clearer. As is also visualized schematically in the zoomed-in region of \figref{fig:secondmodel}a, the unprimed (primed) couplings couple the $d_{i, \alpha}^{\pdagger}$ ($d_{i,\alpha }^{\prime}$) fermions located on different sites. So we see that both of these couplings are required to stabilize a finite bandwidth of both fermion flavors and, thus, to remove an otherwise extensive ground-state degeneracy. Additionally, a finite  $J_{7}$ coupling hybridizes the two mobile Majorana fermion bands. Upon combining the Majorana fermions $d_{i,\alpha }$ and $d_{i,\alpha}^{\prime}$  into complex (Abrikosov) fermions, this term maps onto a chemical potential, thereby explicitly breaking particle-hole symmetry. 

Once again, the bond operators $\hat u_{i_1,i_2}^{{\mu_1,\mu_2}}$ are constants of motion: each can have an eigenvalue $\pm1$ and commutes with the Hamiltonian, so they form a static 
$\mathbb{Z}_2$
gauge field background for the mobile Majorana fermions. The relevant conserved fluxes for the  elementary plaquettes (empty and crossed squares, and triangles inside the latter) are given by the Wilson loop operators $\hat W_{\text{EP}} = \prod_{\langle ij\rangle \in \text{EP}} \hat u_{ij}$, where $\text{EP} \in \{\emptysquare[45],\squarecross[45],  \squaretriangle[45]{1}, \squaretriangle[45]{2},
\squaretriangle[45]{4},\squaretriangle[45]{3}\}$ and all paths are taken in a clockwise way. 

After diagonalizing \equref{eq:spin7/2modelmajor} numerically in momentum space for all unique flux configurations modulo symmetries, we have found four phases in the phase diagram (\figref{fig:secondmodel}b) for fixed $J_{1}=J_{1}^{\prime}=1$, $|J_{2}|\le 0.8$ and $|J_{7}|\le 1.5$. These can be characterized through the six distinct fluxes  on the elementary plaquettes $\left\{\emptysquare[45],\squarecross[45] \mid \squaretriangle[45]{1}, \squaretriangle[45]{2},
\squaretriangle[45]{4},\squaretriangle[45]{3}\right\}_{\vec{R}}$ associated with each unit cell $\vec{R}$ (see \appref{app:furtherdetails} for more details). By considering a super unit cell of eight sites, we visually represent the fluxes of these phases in \figref{fig:secondmodel}c.

Starting from finite $J_{1}$ and $J_{1}^{\prime}$ while $J_{2}=J_{2}^{\prime}=J_{7}=0$, the ground state exhibits a uniform $\pi$ flux on the square $\emptysquare[45]$ plaquettes, consistent with Lieb's theorem \cite{Lieb1989Mar}. Upon introducing the on-site $J_{7}$ couplings, this uniform flux configuration in the square plaquettes persists up to $J_{7}\approx 1.2$. Beyond this threshold, the ground state transitions to a staggered flux pattern with $0$ and $\pi$ fluxes on the $\emptysquare[45]$ and $\squarecross[45]$ plaquettes, respectively. The phases that emerge at small $J_2$ and $J_2^{\prime}$ can be understood as natural extensions of these two flux configurations, now featuring additional non-trivial orbital textures on the triangular elementary plaquettes.

At small $J_{\{2,7\}}$, we find a state with a uniform flux configuration $\left\{-,-\mid -,-,-,-\right\}_{\vec{R}}$, which preserves the rotational symmetry of the lattice but is odd under $\sigma_d$. There is a finite (orbital) magnetic moment in the unit cell and the state represents a chiral spin liquid (denoted by CSL-I in \figref{fig:secondmodel}b). 
Increasing either $J_{2}$ or $J_{7}$ induces more non-trivial orbital textures over the triangle fluxes. First, we obtain a state where the orbital textures alternate in an antiferromagnetic fashion along one direction ($\hat{x}$ in \figref{fig:secondmodel}c); we refer to this state as an orbital stripy spin liquid (OSSL). Most importantly for us here, for larger $J_7$, we again obtain an OAMSL: as can be seen from the flux configurations depicted in \figref{fig:secondmodel}c, it preserves translations, $\sigma_d$, and the combination of time-reversal $\Theta$ and $C_{4z}$, while those last two symmetries are broken individually. As such the associated altermagnetic order parameter transforms under $B_2$ of $C_{4v}$; it can, thus, be thought of as a $d$-wave OAMSL and is very closely related to the state discussed in \refcite{Sobral2025May}. We note for completeness that being even under $\sigma_d$ is already sufficient to force the magnetization to vanish since the only finite moments are orbital moments, which are Ising-like in two spatial dimensions.

Finally, these orbital textures are destroyed upon increasing $J_{7}$ further, giving rise to a second chiral spin liquid (CSL-II) with a checkerboard flux pattern between elementary square plaquettes $\left\{-,+\mid -,-,-,-\right\}_{\vec{R}}$. 

By considering $|J_{\{2,7\}}|\ge 1.5$, we found an even richer phase diagram with additional phases that can be stabilized, including $p$-wave OAMSLs, as well as regions of increased degeneracy. Apart from CSL-I and CSL-II, all these phases were found to have gapless spinon excitations for the considered couplings $J_{\{2,7\}}$. We refer the reader to \figref{fig:extendedphasediagram72} and \appref{app:furtherdetails} for more details.

\section{Flux excitations for the OAMSL}

\begin{figure}[t]
    \centering
    \includegraphics[width=1\linewidth]{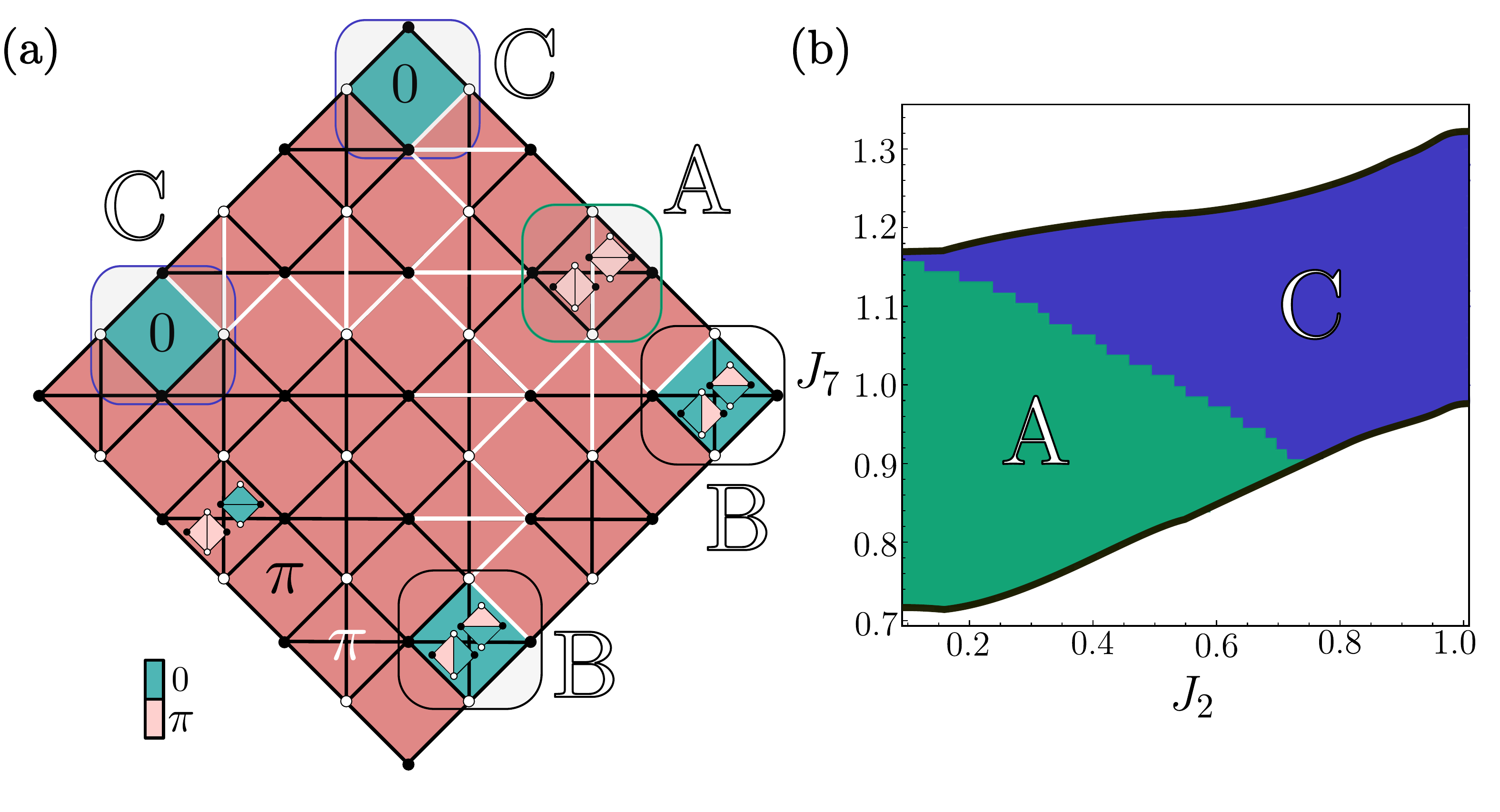}
    \caption{(a) Schematic representation of flux excitations in the OAMSL phase of the spin-7/2 model. 
    The non-topological excitation (A) consists of localized flux flips within a single plaquette, requiring no gauge string for creation or annihilation. Two topological excitations (visons B and C) are shown with their characteristic gauge strings (white); the gauge strings carry zero energy cost and enable free vison propagation. We indicate the modified flux patterns on triangular plaquettes relative to the staggered ground state configuration (not shown for clarity). (b) Type of the lowest flux excitation within the OAMSL domain as a function of $J_{2}$ and $J_{7}$ for $J_{2}^{\prime}=J_{2}$ and $J_{1}^{\prime}=J_{1}=1$. See also Fig.~\ref{fig:supvison} for more details.}
    \label{fig:visons}
\end{figure}

Having established the ground state properties of the spin models, we now examine their flux excitations. Our primary focus is the OAMSL in the spin-7/2 model, though this analysis can be readily extended to other phases, including those in the spin-3/2 model. There are three possible elementary excitations in the flux sector: one non-topological (A) and two topological (visons B and C), as illustrated in \figref{fig:visons}a. 

The non-topological excitations (A) arise from bond flips within filled squares \squarecross[45]. These excitations manifest as local ferromagnetic moments, characterized by uniform fluxes inside the elementary triangles of the lattice, i.e., $(+,+,+,+)$ or its time-reversal conjugate $(-,-,-,-)$. This locally breaks $\sigma_d$ and thus can be thought of as the local creation of a chiral configuration within a single unit cell. Since they can be created within a single filled square and moved without requiring a gauge string, they lack the topological character of visons.
 
As for the topological excitations, we first consider configurations where empty square fluxes remain fixed, while triangles and filled square \squarecross[45] fluxes undergo minimal changes. This corresponds to vison-B excitations with dipolar endpoints that break the original alternating $(+,-,+,-)$ 
flux pattern of triangles in the  OAMSL. At the string endpoints, the flux configurations become $(-,-,+,+)$ and $(-,+,+,-)$ configurations leading to local dipolar structures---in contrast to the quadrupolar structure of the underlying OAMSL.
The endpoints can propagate independently at opposite ends of the string but necessarily leave a string of flipped gauge configurations behind, signaling their topological vison-like nature. 
Moreover, each endpoint's dipole can independently rotate to $(+,+,-,-)$, $(-,+,+,-)$, $(-,-,+,+)$ or $(+,-,-,+)$. This yields four distinct states per endpoint. Consequently, these defects carry additional quantum numbers compared to conventional visons in simpler $\mathbb{Z}_2$ spin liquids, a direct consequence of the sublattice structure. These excitations can move freely and constitute genuine topological excitations that, once separated, cannot be annihilated through local operations.

If we also allow the empty square plaquettes to flip, we obtain a second type of vison excitation, labeled by C in \figref{fig:visons}a. They do not break any symmetries and, as opposed to the vison-B excitations, do not carry additional internal quantum numbers. 

Figure \ref{fig:visons}b indicates in the parameter regime of the OAMSL which excitation type has the lowest energy. The non-topological excitation A exhibits the smallest gap at low $J_2$ coupling, while vison-C becomes preferred as $J_2$ increases. 
Despite vison-B appearing to have a larger gap throughout this domain, we note a region in parameter space near $J_7 \approx 1.1$ at small $J_{2}$ values, where all three types of excitations are nearly degenerate (see \figref{fig:supvison}).

\section{Conclusion}
We have constructed and analyzed two exactly solvable two-dimensional spin models hosting orbital altermagnetic and other competing quantum spin liquids. Although both are defined on tetragonal Bravais lattices, their phase diagrams differ significantly: the first model can be thought of as spin-$3/2$ degrees of freedom (or two qubits) on the sites of a square-octagon lattice, see \figref{fig:firstmodel}(a) or \equref{eq:spin3/2model}. Despite the large number of intra-unit cell degrees of freedom and the fact that the lattice geometry allows for various other orbital textures, we found that a $g$-wave orbital altermagnetic spin liquid consistently exhibits the lowest ground-state energy in the parameter regime we have considered. The second model we constructed is defined on the checkerboard lattice, which requires three qubits per site (or spin-7/2) to retain exact solvability via a Majorana representation. As opposed to the spin-$3/2$ model, the phase diagram is rich and features several distinct spin liquids, including chiral spin liquids, translational-symmetry breaking spin liquids, but also a $d$-wave altermagnetic spin liquid. We further analyzed the possible excitations in the flux sector, which are summarized graphically in \figref{fig:visons}a and reveal both topological vison-like excitations that are associated with gauge strings (B and C) as well as non-topological local ferromagnetic flips of the gauge fields (A).
In the future, it would be interesting to explore the impact of additional perturbations, such as a static magnetic fields in analogy with the Kitaev model \cite{KITAEV20062} or higher-order interactions \cite{Zhang2019Jul, Takikawa2020Nov}. Another import future direction would be to analyze how these models and, hence, orbital altermagnetic phases in general, could be stabilized in programmable quantum-simulation platforms, such as Rydberg-atom arrays \cite{rydberg}.

\begin{acknowledgments}
The authors thank Vitor Dantas, Rodrigo G. Pereira, and Stefano Chesi for insightful discussions.  
S.M.~and M.S.S.~further acknowledge funding by the European Union (ERC-2021-STG, Project 101040651---SuperCorr). Views and opinions expressed are however those of the authors only and do not necessarily reflect those of the European Union or the European Research Council Executive Agency. Neither the European Union nor the granting authority can be held responsible for them.
P.M.B. acknowledges support by the German National Academy of Sciences Leopoldina through Grant No.~LPDS 2023-06, by the Gordon and Betty Moore Foundation’s EPiQS Initiative Grant GBMF8683, and by the NSF Grant DMR-2245246.
\end{acknowledgments}

\bibliography{draft_Refs}

@article{Vijayvargia2025Oct,
	author = {Vijayvargia, Aayush and Day-Roberts, Ezra and Botana, Antia S. and Erten, Onur},
	title = {{Altermagnets with Topological Order in Kitaev Bilayers}},
	journal = {Phys. Rev. Lett.},
	volume = {135},
	number = {16},
	pages = {166701},
	year = {2025},
	month = oct,
	publisher = {American Physical Society},
	doi = {10.1103/km2j-3zy2}
}

@article{Yao2009May,
	author = {Yao, Hong and Zhang, Shou-Cheng and Kivelson, Steven A.},
	title = {{Algebraic Spin Liquid in an Exactly Solvable Spin Model}},
	journal = {Phys Rev Lett},
	volume = {102},
	number = {21},
	pages = {217202},
	year = {2009},
	month = may,
	publisher = {American Physical Society},
	doi = {10.1103/PhysRevLett.102.217202}
}

@article{Pedrocchi2011Oct,
	author = {Pedrocchi, Fabio L. and Chesi, Stefano and Loss, Daniel},
	title = {{Physical solutions of the Kitaev honeycomb model}},
	journal = {Phys Rev B},
	volume = {84},
	number = {16},
	pages = {165414},
	year = {2011},
	month = oct,
	publisher = {American Physical Society},
	doi = {10.1103/PhysRevB.84.165414}
}

@article{Zschocke2015Jul,
	author = {Zschocke, Fabian and Vojta, Matthias},
	title = {{Physical states and finite-size effects in Kitaev's honeycomb model: Bond disorder, spin excitations, and NMR line shape}},
	journal = {Phys Rev B},
	volume = {92},
	number = {1},
	pages = {014403},
	year = {2015},
	month = jul,
	publisher = {American Physical Society},
	doi = {10.1103/PhysRevB.92.014403}
}

@article{Chua2011May,
	author = {Chua, Victor and Yao, Hong and Fiete, Gregory A.},
	title = {{Exact chiral spin liquid with stable spin Fermi surface on the kagome lattice}},
	journal = {Phys Rev B},
	volume = {83},
	number = {18},
	pages = {180412},
	year = {2011},
	month = may,
	publisher = {American Physical Society},
	doi = {10.1103/PhysRevB.83.180412}
}

@article{rydberg,
  title = {Proposal for realization and detection of Kitaev quantum spin liquid with Rydberg atoms},
  author = {Chen, Yi-Hong and Wang, Bao-Zong and Poon, Ting-Fung Jeffrey and Zhou, Xin-Chi and Liu, Zheng-Xin and Liu, Xiong-Jun},
  journal = {Phys. Rev. Res.},
  volume = {6},
  issue = {4},
  pages = {L042054},
  numpages = {7},
  year = {2024},
  month = {Dec},
  publisher = {American Physical Society},
  doi = {10.1103/PhysRevResearch.6.L042054},
  url = {https://link.aps.org/doi/10.1103/PhysRevResearch.6.L042054}
}

@article{smejkal_emerging_2022,
  title = {Emerging {{Research Landscape}} of {{Altermagnetism}}},
  author = {{\v S}mejkal, Libor and Sinova, Jairo and Jungwirth, Tomas},
  year = {2022},
  month = dec,
  journal = {Physical Review X},
  volume = {12},
  number = {4},
  pages = {040501},
  issn = {2160-3308},
  doi = {10.1103/PhysRevX.12.040501},
  urldate = {2024-01-09},
  langid = {english},
}

@article{Zhou2017Apr,
	author = {Zhou, Yi and Kanoda, Kazushi and Ng, Tai-Kai},
	title = {{Quantum spin liquid states}},
	journal = {Rev. Mod. Phys.},
	volume = {89},
	number = {2},
	pages = {025003},
	year = {2017},
	month = apr,
	publisher = {American Physical Society},
	doi = {10.1103/RevModPhys.89.025003}
}

@article{Vijayvargia2023Jun,
	author = {Vijayvargia, Aayush and Nica, Emilian Marius and Moessner, Roderich and Lu, Yuan-Ming and Erten, Onur},
	title = {{Magnetic fragmentation and fractionalized Goldstone modes in a bilayer quantum spin liquid}},
	journal = {Phys. Rev. Res.},
	volume = {5},
	number = {2},
	pages = {L022062},
	year = {2023},
	month = jun,
	publisher = {American Physical Society},
	doi = {10.1103/PhysRevResearch.5.L022062}
}

@article{Petit2016Aug,
	author = {Petit, S. and Lhotel, E. and Canals, B. and Ciomaga Hatnean, M. and Ollivier, J. and Mutka, H. and Ressouche, E. and Wildes, A. R. and Lees, M. R. and Balakrishnan, G.},
	title = {{Observation of magnetic fragmentation in spin ice}},
	journal = {Nat. Phys.},
	volume = {12},
	pages = {746--750},
	year = {2016},
	month = aug,
	issn = {1745-2481},
	publisher = {Nature Publishing Group},
	doi = {10.1038/nphys3710}
}

@article{Brooks-Bartlett2014Jan,
	author = {Brooks-Bartlett, M. E. and Banks, S. T. and Jaubert, L. D. C. and Harman-Clarke, A. and Holdsworth, P. C. W.},
	title = {{Magnetic-Moment Fragmentation and Monopole Crystallization}},
	journal = {Phys. Rev. X},
	volume = {4},
	number = {1},
	pages = {011007},
	year = {2014},
	month = jan,
	publisher = {American Physical Society},
	doi = {10.1103/PhysRevX.4.011007}
}

@article{yu2024altermagnetism,
	author = {Yu, Yue and Suh, Han Gyeol and Roig, Merc{\ifmmode\grave{e}\else\`{e}\fi} and Agterberg, Daniel F.},
	title = {{Altermagnetism from coincident Van Hove singularities: application to {$\kappa$}-Cl}},
	journal = {Nat Commun},
	volume = {16},
	number = {2950},
	pages = {1--8},
	year = {2025},
	month = mar,
	issn = {2041-1723},
	publisher = {Nature Publishing Group},
	doi = {10.1038/s41467-025-57970-9}
}

@article{banerjeeAltermagneticSuperconductingDiode2024,
  title = {Altermagnetic superconducting diode effect},
  author = {Banerjee, Sayan and Scheurer, Mathias S.},
  journal = {Phys. Rev. B},
  volume = {110},
  issue = {2},
  pages = {024503},
  numpages = {13},
  year = {2024},
  month = {Jul},
  publisher = {American Physical Society},
  doi = {10.1103/PhysRevB.110.024503},
  url = {https://link.aps.org/doi/10.1103/PhysRevB.110.024503}
}

@article{Wen2002Apr,
	author = {Wen, Xiao-Gang},
	title = {{Quantum orders and symmetric spin liquids}},
	journal = {Phys. Rev. B},
	volume = {65},
	number = {16},
	pages = {165113},
	year = {2002},
	month = apr,
	publisher = {American Physical Society},
	doi = {10.1103/PhysRevB.65.165113}
}

@article{Nayak2000Aug,
	author = {Nayak, Chetan},
	title = {{Density-wave states of nonzero angular momentum}},
	journal = {Phys. Rev. B},
	volume = {62},
	number = {8},
	pages = {4880--4889},
	year = {2000},
	month = aug,
	publisher = {American Physical Society},
	doi = {10.1103/PhysRevB.62.4880}
}

@article{Senthil2004Jan,
	author = {Senthil, T. and Vojta, Matthias and Sachdev, Subir},
	title = {{Weak magnetism and non-Fermi liquids near heavy-fermion critical points}},
	journal = {Phys. Rev. B},
	volume = {69},
	number = {3},
	pages = {035111},
	year = {2004},
	month = jan,
	publisher = {American Physical Society},
	doi = {10.1103/PhysRevB.69.035111}
}

@article{Zhang2019Jul,
	author = {Zhang, Shang-Shun and Wang, Zhentao and Hal{\ifmmode\acute{a}\else\'{a}\fi}sz, G{\ifmmode\acute{a}\else\'{a}\fi}bor B. and Batista, Cristian D.},
	title = {{Vison Crystals in an Extended Kitaev Model on the Honeycomb Lattice}},
	journal = {Phys. Rev. Lett.},
	volume = {123},
	number = {5},
	pages = {057201},
	year = {2019},
	month = jul,
	publisher = {American Physical Society},
	doi = {10.1103/PhysRevLett.123.057201}
}

@article{Lieb1989Mar,
	author = {Lieb, Elliott H.},
	title = {{Two theorems on the Hubbard model}},
	journal = {Phys. Rev. Lett.},
	volume = {62},
	number = {10},
	pages = {1201--1204},
	year = {1989},
	month = mar,
	publisher = {American Physical Society},
	doi = {10.1103/PhysRevLett.62.1201}
}

@article{Jungwirth2025Aug,
	author = {Jungwirth, Tom{\ifmmode\acute{a}\else\'{a}\fi}{\ifmmode\check{s}\else\v{s}\fi} and Fernandes, Rafael M. and Fradkin, Eduardo and MacDonald, Allan H. and Sinova, Jairo and {\ifmmode\check{S}\else\v{S}\fi}mejkal, Libor},
	title = {{Altermagnetism: An unconventional spin-ordered phase of matter}},
	journal = {Newton},
	volume = {1},
	number = {6},
	year = {2025},
	month = aug,
	issn = {2950-6360},
	publisher = {Elsevier},
	doi = {10.1016/j.newton.2025.100162}
}

@article{Fernandes2024Jan,
	author = {Fernandes, Rafael M. and de Carvalho, Vanuildo S. and Birol, Turan and Pereira, Rodrigo G.},
	title = {{Topological transition from nodal to nodeless Zeeman splitting in altermagnets}},
	journal = {Phys. Rev. B},
	volume = {109},
	number = {2},
	pages = {024404},
	year = {2024},
	month = jan,
	publisher = {American Physical Society},
	doi = {10.1103/PhysRevB.109.024404}
}

@article{Takikawa2020Nov,
	author = {Takikawa, Daichi and Fujimoto, Satoshi},
	title = {{Topological phase transition to Abelian anyon phases due to off-diagonal exchange interaction in the Kitaev spin liquid state}},
	journal = {Phys. Rev. B},
	volume = {102},
	number = {17},
	pages = {174414},
	year = {2020},
	month = nov,
	publisher = {American Physical Society},
	doi = {10.1103/PhysRevB.102.174414}
}

@article{Demler1999Jan,
	author = {Demler, Eugene and Zhang, Shou-Cheng},
	title = {{Non-Abelian Holonomy of BCS and SDW Quasiparticles}},
	journal = {Ann. Phys.},
	volume = {271},
	number = {1},
	pages = {83--119},
	year = {1999},
	month = jan,
	issn = {0003-4916},
	publisher = {Academic Press},
	doi = {10.1006/aphy.1998.5866}
}

@article{Ghaemi2006Feb,
	author = {Ghaemi, Pouyan and Senthil, T.},
	title = {{N{\ifmmode\backslash\else\textbackslash\fi}'eel order, quantum spin liquids, and quantum criticality in two dimensions}},
	journal = {Phys. Rev. B},
	volume = {73},
	number = {5},
	pages = {054415},
	year = {2006},
	month = feb,
	publisher = {American Physical Society},
	doi = {10.1103/PhysRevB.73.054415}
}

@article{Liu2025Oct,
	author = {Liu, Zhao and Hu, Hui and Liu, Xia-Ji},
	title = {{Altermagnetism and Superconductivity: A Short Historical Review}},
	journal = {arXiv},
	year = {2025},
	month = oct,
	eprint = {2510.09170},
	doi = {10.48550/arXiv.2510.09170}
}

@article{Neehus2025Apr,
  title = {Projectively Implemented Altermagnetism in an Exactly Solvable Quantum Spin Liquid},
  author = {Neehus, Avedis and Rosch, Achim and Knolle, Johannes and Seifert, Urban F. P.},
  journal = {Phys. Rev. Lett.},
  volume = {135},
  issue = {25},
  pages = {256504},
  numpages = {7},
  year = {2025},
  month = {Dec},
  publisher = {American Physical Society},
  doi = {10.1103/cnk8-vnxg},
  url = {https://link.aps.org/doi/10.1103/cnk8-vnxg}
}

@article{PhysRevLett.134.146001,
  title = {Adatom Engineering Magnetic Order in Superconductors: Applications to Altermagnetic Superconductivity},
  author = {Pupim, Lucas V. and Scheurer, Mathias S.},
  journal = {Phys. Rev. Lett.},
  volume = {134},
  issue = {14},
  pages = {146001},
  numpages = {9},
  year = {2025},
  month = {Apr},
  publisher = {American Physical Society},
  doi = {10.1103/PhysRevLett.134.146001},
  url = {https://link.aps.org/doi/10.1103/PhysRevLett.134.146001}
}

@ARTICLE{2025arXiv251000509P,
       author = {{Pan}, Mingxiang and {Liu}, Feng and {Huang}, Huaqing},
        title = "{Orbital Altermagnetism}",
      journal = {arXiv e-prints},
         year = 2025,
        month = oct,
archivePrefix = {arXiv},
       eprint = {2510.00509},
 primaryClass = {cond-mat.mtrl-sci},
       adsurl = {https://ui.adsabs.harvard.edu/abs/2025arXiv251000509P}
}

@ARTICLE{2025arXiv250926596C,
       author = {{Chakraborty}, Anzumaan R. and {Yang}, Fan and {Birol}, Turan and {Fernandes}, Rafael M.},
        title = "{Orbital altermagnetism on the kagome lattice and possible application to $A$V$_3$Sb$_5$}",
      journal = {arXiv e-prints},
         year = 2025,
        month = sep,
archivePrefix = {arXiv},
       eprint = {2509.26596},
 primaryClass = {cond-mat.str-el},
       adsurl = {https://ui.adsabs.harvard.edu/abs/2025arXiv250926596C}
}

@article{Savary_2017,
doi = {10.1088/0034-4885/80/1/016502},
url = {https://dx.doi.org/10.1088/0034-4885/80/1/016502},
year = {2016},
month = {nov},
publisher = {IOP Publishing},
volume = {80},
number = {1},
pages = {016502},
author = {Savary, Lucile and Balents, Leon},
title = {Quantum spin liquids: a review},
journal = {Reports on Progress in Physics},
}

@article{KITAEV20062,
title = {Anyons in an exactly solved model and beyond},
journal = {Annals of Physics},
volume = {321},
number = {1},
pages = {2-111},
year = {2006},
note = {January Special Issue},
issn = {0003-4916},
doi = {https://doi.org/10.1016/j.aop.2005.10.005},
url = {https://www.sciencedirect.com/science/article/pii/S0003491605002381},
author = {Alexei Kitaev},
}

@article{Sobral2025May,
	author = {Sobral, Jo{\ifmmode\tilde{a}\else\~{a}\fi}o Augusto and Mandal, Subrata and Scheurer, Mathias S.},
	title = {{Fractionalized altermagnets: From neighboring and altermagnetic spin liquids to spin-symmetric band splitting}},
	journal = {Phys. Rev. Res.},
	volume = {7},
	number = {2},
	pages = {023152},
	year = {2025},
	month = may,
	publisher = {American Physical Society},
	doi = {10.1103/PhysRevResearch.7.023152}
}

@article{PhysRevLett.99.247203,
  title = {Exact Chiral Spin Liquid with Non-Abelian Anyons},
  author = {Yao, Hong and Kivelson, Steven A.},
  journal = {Phys. Rev. Lett.},
  volume = {99},
  issue = {24},
  pages = {247203},
  numpages = {4},
  year = {2007},
  month = {Dec},
  publisher = {American Physical Society},
  doi = {10.1103/PhysRevLett.99.247203},
  url = {https://link.aps.org/doi/10.1103/PhysRevLett.99.247203}
}

@article{Murakami,
  title = {$\text{SU}(2)$ non-Abelian holonomy and dissipationless spin current in semiconductors},
  author = {Murakami, Shuichi and Nagosa, Naoto and Zhang, Shou-Cheng},
  journal = {Phys. Rev. B},
  volume = {69},
  issue = {23},
  pages = {235206},
  numpages = {14},
  year = {2004},
  month = {Jun},
  publisher = {American Physical Society},
  doi = {10.1103/PhysRevB.69.235206},
  url = {https://link.aps.org/doi/10.1103/PhysRevB.69.235206}
}

@article{PhysicalsolKSL,
  title = {Physical solutions of the Kitaev honeycomb model},
  author = {Pedrocchi, Fabio L. and Chesi, Stefano and Loss, Daniel},
  journal = {Phys. Rev. B},
  volume = {84},
  issue = {16},
  pages = {165414},
  numpages = {7},
  year = {2011},
  month = {Oct},
  publisher = {American Physical Society},
  doi = {10.1103/PhysRevB.84.165414},
  url = {https://link.aps.org/doi/10.1103/PhysRevB.84.165414}
}

\clearpage
\onecolumngrid
\appendix

\addcontentsline{toc}{section}{Supplementary Material}

\begin{center}
\textbf{Supplementary Material for 
Exactly Solvable Models Hosting Altermagnetic Quantum Spin Liquids}

\vspace{0.5em}
João Augusto Sobral$^{1}$, Pietro M. Bonetti$^{2}$, Subrata Mandal$^{1}$, Mathias S.~Scheurer$^{1}$

\vspace{0.3em}
{\small
$^{1}$Institute for Theoretical Physics III, University of Stuttgart, Germany\\
$^{2}$Department of Physics, Harvard University, USA
}
\end{center}

\begin{appendix}

\section{Local physical constraints for Majoranas}
\label{app:physicalstates}

Kitaev's solution of the honeycomb-lattice model offers a significant simplification by mapping the diagonalization of a $2^{2N} \times 2^{2N}$ Hamiltonian to the problem of diagonalizing $2N \times 2N$ matrix \cite{KITAEV20062}. However, this mapping comes at the expense of enlarging the Hilbert space through the introduction of Majorana fermions; thereby, the resulting eigenstates of the Majorana Hamiltonian may not correspond to physical states of the original spin system. To recover the physical ground state, a projection onto the physical subspace is necessary. This can be enforced by the local gauge constraints $D_i | \psi \rangle = | \psi \rangle$ at each site $i$. To be more precise, it can be shown that  the operators $D_i$ generate the projection operator $P$, which can be written explicitly as $P=\mathcal{S}P_0$, where $\mathcal{S}$ contains all inequivalent gauge transformations and $P_0=(1+D)/2$ with $D \equiv \prod_{i} D_i$.

\subsection{The projection operator in the  square-octagon lattice}
\begin{figure}[h]
    \centering
    \includegraphics[width=1\linewidth]{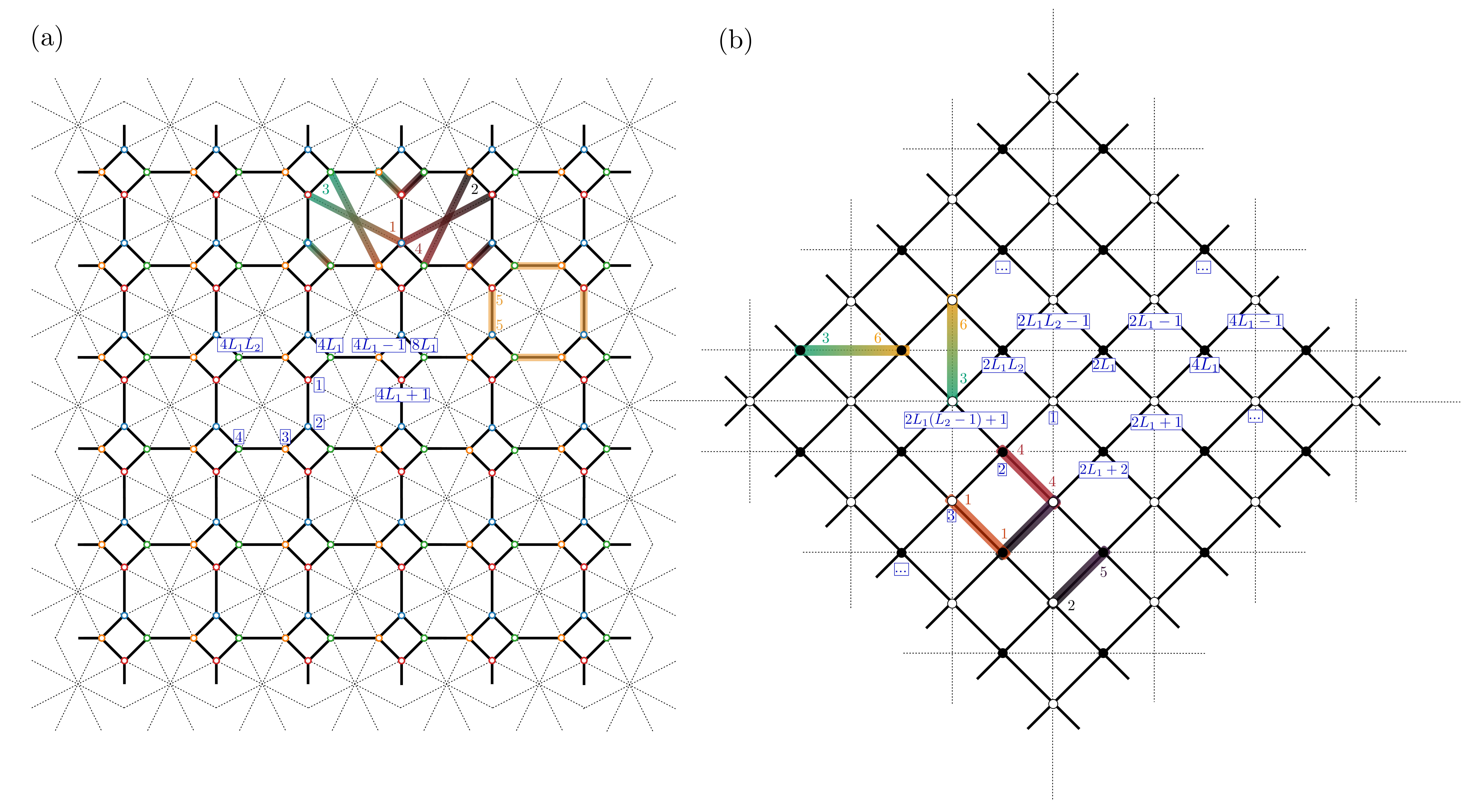}
            \caption{Boundary conditions for the spin-3/2 (a) and spin-7/2 models (b). Each site is indicated through blue boxed labels. The representative bonds $\hat u_{ij}^{\alpha,\tau}$ for each model are represented and colored in accordance with \figref{fig:firstmodel}a and \figref{fig:secondmodel}a.}
    \label{fig:bc}
\end{figure}

This appendix presents the rewriting of the projection operator in the  square-octagon lattice in terms of fermion parities and link variables. More specifically, we focus on a square-octagon lattice embedded on a torus, with the labeling shown in Fig. \ref{fig:bc}a. We begin with the definition:
\begin{equation}
2P^{(3/2)}_0 = 1 + D^{(3/2)},
\end{equation}
where the gauge operator is defined as $D^{(3/2)}=\prod_{i} D^{(3/2)}_i = \prod_{i} (-i) c^1_i c^2_i c^3_i c^4_i c^5_i d_i =  \prod_{i} c^1_i c^2_i c^3_i c^4_i c^5_i d_i$ . Here, we also assume that the total number of sites in our square-octagonal lattice is $N = 4L_1L_2$.

To reexpress the projection operator in terms of the link operators, first, we bring all $c$-Majorana operators to the right of the $d$-Majorana operators. Next, we group the $c^1$ and $c^3$ operators together, as well as the $c^2$ and $c^4$ operators, so that they can be associated with $u$-operators. When we rearrange the operators, the anticommutation of the Majorana operator produces an overall factor of $(-1)^N$. Since $N$ is even in our case, this phase factor becomes unity, and the regrouping can be carried out without additional sign. This leads us to  
\begin{equation}
D^{(3/2)} = \prod^N_{i} c^1_ic^3_i \prod^N_{j} c^2_jc^4_j \prod^N_{k} c^5_kd_k .
\end{equation}
The remaining product can be further simplified by separating the operators $c^5$ and $d$. This can again be done without introducing any phase factor. Thus, operator $D$ decomposes into the product of four independent pieces, yielding
\begin{equation}
D^{(3/2)} = \prod^N_{i} c^1_ic^3_i \prod^N_{j} c^2_jc^4_j \prod^N_{k} c^5_k \prod^N_{l}d_l .
\end{equation}
When we evaluate the first three pieces involving the $c$-operators, they reduce to the product of $u$-operators, accompanied by some additional phase factors that originate from the rearrangement of Majorana operators. After simplification, these products turn out to be 
\begin{align}
& \prod^N_{i} c^1_ic^3_i  = (-1)^{L_1L_2} \prod_{ij} u^{31}_{ij}, \\
& \prod^N_{i} c^2_ic^4_i  = (-1)^{L_1L_2}  \prod_{ij} u^{24}_{ij}, \\
& \prod^N_{i} c^5_i   = (-1)^{L_1L_2} \prod_{ij} u^{55}_{ij},
\end{align}
We are now only left with $d$-Majorana operators. To evaluate the contribution of these $d$-Majorana operators, it is convenient to introduce complex bond fermions, which are defined as $f_1=(d_C-id_A)/2$ and $f_2=(d_B-id_D)/2$. As a result, within this representation, the product over $d$-Majoranas can be expressed as 
\begin{equation}
\prod d_i = (-1)^{L_1L_2}(-1)^{N_{\chi}},
\end{equation}
where $N_{\chi}$ denotes the number of bond fermions.

This leads directly to the explicit form of the projection operator given by
\begin{equation}
2P^{(3/2)}_0 =1+  (-1)^{N_{\chi}} \prod u^{31}_{ij} \prod u^{55}_{kl} \prod u^{24}_{mn}. 
\end{equation}

\subsection{The projection operator in the checkerboard lattice}
In this section, we derive the projection operator for the checkerboard lattice, which is embedded on a torus, and adopt the labelling illustrated in Fig. \ref{fig:bc}b. We again start from the definition 
\begin{equation}
2P^{(7/2)}_0 = 1 + D^{(7/2)},
\end{equation}
but with $D^{(7/2)}=\prod_{i} D^{(7/2)}_i= \prod_{i} c^1_i c^2_i c^3_i c^4_i c^5_i c^6_i  d_i d'_i $ as the gauge operator. Here, we consider the lattice comprises $2L_1 L_2$ sites. 
In the first step, we move all $d$ and $d'$ operators to the left of the $c$-operators, and we reorder the $c$-operators so that $c^1$ and $c^4$, $c^6$ and $c^3$, and $c^2$ and $c^5$ are adjacent. This rearrangement does not generate any sign change, resulting  
\begin{equation}
D^{(7/2)} = \prod^N_{i} c^1_ic^4_i \prod^N_{j} c^6_jc^3_j \prod^N_{k} c^2_kc^5_k \prod^N_{m} d_m d'_m  .
\end{equation}
At this stage, we separate the $c^1$ and $c^4$ operators, which introduces a phase factor. This yields
\begin{equation}
D^{(7/2)} = (-1)^{{\theta}_1}\prod^N_{i_1} c^1_{i_1}  \prod^N_{i_2}c^4_{i_2} \prod^N_{j} c^6_jc^3_j \prod^N_{k} c^2_kc^5_k \prod^N_{m} d_m d'_m ,
\end{equation}
where, ${\theta}_1 = L_1L_2(2L_1L_2-1)$. Now we are ready to write the $c$-operators in terms of $u$-operators. This step introduces additional phase factors as before. After all the algebraic manipulation, we arrive at  
\begin{align}
& \prod^N_{i} c^6_ic^3_i  = (-1)^{{\theta}_2}(-1)^{L_1L_2} \prod_{ij} u^{63}_{ij}, \\
& \prod^N_{i} c^1_i  =  (-1)^{{\theta}_3}(i)^{L_1L_2} \prod_{ij} u^{11}_{ij}, \\
& \prod^N_{i} c^4_i  =  (-1)^{{\theta}_4}(i)^{L_1L_2} \prod_{ij} u^{44}_{ij}, \\
& \prod^N_{i} c^2_i c^5_i   = (-1)^{L_1L_2+L_2} \prod_{ij} u^{52}_{ij},
\end{align}
where, ${\theta}_2=L_2+L_1L_2$, ${\theta}_3=L_1+L_1L_2$, and ${\theta}_4=L_1+L_2+L_1L_2$. 
Now the remaining Majorana degrees of freedom are $d$ and $d'$  at each site. To proceed further, we introduce a complex fermion at each site, defined as $f_i=(d_i-id'_i)/2$. The product of $d$ and $d'$ Majorana operators can then be written explicitly:
\begin{equation}
\prod  d_i d'_i = (-1)^{L_1L_2} (-1)^{N_{\chi}},
\end{equation}
where, $N_{\chi}$ counts the number of complex fermions at each site.
\\

Hence, the projection operator takes the form
\begin{equation}
2P^{(7/2)}_0 =1+ (-1)^{L_2} (-1)^{N_{\chi}} \prod u^{11}_{ij} \prod u^{44}_{mn} \prod u^{63}_{pq} \prod u^{52}_{uv}. 
\end{equation}

In this section, we derived two projection operators that determine whether a state $|\Psi \rangle $ lies within the physical subspace or outside of it. More importantly, these operators impose a definite fermionic parity, either even or odd, for a state to be physical. The fermionic parity is generally decided by the gauge configuration and the imposed boundary condition. 
Once the gauge configuration that minimizes the energy is obtained, the occupation of the complex fermionic can always be adjusted so that the resulting state corresponds to the true physical ground state. We would also like to emphasize that the explicit form of the projection operator depends on both the boundary conditions and the choice of labeling \cite{PhysicalsolKSL,Zschocke2015Jul}. They altogether dictate the physicality of the ground state.

\section{Symmetries, Clifford generators and irreducible representations of SU(2)}\label{app:higherorder}

\subsection{Representation of symmetries}
The specific representation for the symmetry operators  that transform the Clifford generators in \equsref{eq:spin-3/2map}{eq:spin-7/2map} according to \equsref{eq:sym32}{eq:sym72} are given by
\begin{align}
\label{eq:sym7/2rep}
C_{4z} &= U_{\pi}\left(4,5\right) U_{\pi}\left(1,2\right) U_{\pi/2}\left(1,4\right)U_{\pi/2}\left(2,4\right) U_{\pi/2}\left(3,4\right),  \\ 
\sigma_v &= U_{\pi}\left(2,4\right) U_{\pi/2}\left(3,4\right)U_{\pi/2}\left(1,2\right),  \\
C_{2z} &= \mathcal{U}_{\pi/2}\left(4,1\right) \mathcal{U}_{\pi/2}\left(5,2\right) \mathcal{U}_{\pi}\left(5,7\right) \mathcal{U}_{\pi/2}\left(6,3\right) \mathcal{U}_{\pi}\left(6,7\right), \\   
\sigma_d&=\mathbb{I}_8,
\end{align}
where the unitaries are defined as $U_\theta=\exp(\frac{\theta}{4}\left[\Gamma^{a},\Gamma^{b}\right])$ and $\mathcal{U}_\omega=\exp(\frac{\omega}{4}\left[\gamma^{a},\gamma^{b}\right])$.  The anti-unitary operator $\Theta$ that transforms the $\Gamma^{a}$-matrices in \equref{eq:spin-3/2map} as $\Theta_{3/2}^{\pdagger} \Gamma^{a} \Theta_{3/2} ^{\dagger}=\Gamma^{a}$ is represented as $\Theta_{3/2} = \sigma_{x} \otimes \sigma_y \mathcal{K}$, where $\mathcal{K}$ stands for complex conjugation. 

In contrast for the spin-$7/2$ model, not all $\gamma^{a}$-matrices can be even under time-reversal. We choose $\Theta_{7/2} = i \sigma_{y} \otimes\sigma_{x} \otimes \mathbb{I}_2 \mathcal{K}$ such that $\Theta_{7/2}^{\pdagger} \gamma^a \Theta_{7/2}^\dagger = s_a\gamma^a$ with $s_1=s_7=+1$ and $s_2=s_3=s_4=s_5=s_6=-1$. This is clearly a symmetry of $\mathcal{H}_{7/2}$ in \equref{eq:spin7/2model}. Modulo $\mathbb{Z}_2$ gauge transformations, the Majorana particles transform as
\begin{eqnarray}
    \Theta_{7/2}^{\pdagger} d_i \Theta_{7/2}^{\dagger} = -d_i, \quad \Theta_{7/2}^{\pdagger} d'_i \Theta_{7/2}^{\dagger} = d'_i, \quad \Theta_{7/2}^{\pdagger} c^a_i \Theta_{7/2}^{\dagger} = s_a c^a_i,
\end{eqnarray}
which is consistent with the behavior of \equref{eq:majorana7/2} under time-reversal. Consequently, the link variables transform as $\hat{u}_{i_1,i_2}^{a_1,a_2} = i c_{i_1}^{a_1} c_{i_2}^{a_2} \rightarrow - s_{a_1} s_{a_2} \hat{u}_{i_1,i_2}^{a_1,a_2}$. However, note that only the combinations $(a_1,a_2) \in \{(1,1),(4,4),(2,5),(3,6)\}$ matter (cf.~\figref{fig:secondmodel}a), for which $s_{a_1} s_{a_2} = 1$. Consequently, we effectively again obtain $\hat{u}_{i_1,i_2}^{a_1,a_2} \rightarrow -\hat{u}_{i_1,i_2}^{a_1,a_2}$ under time-reversal, just like in the spin-3/2 model.

Finally, we point out that $\sigma_v$ can be represented in a sublattice dependent way. Equation~(\ref{eq:spin7/2model}) stays invariant under $\sigma_v$ if
\begin{subequations}\begin{align}
   \text{sublattice A:} \quad &(\gamma^1, \gamma^2) \leftrightarrow (\gamma^5, \gamma^4), \, \gamma_{3,6,7} \mapsto \gamma_{3,6,7} \\
   \text{sublattice B:} \quad &(\gamma^1, \gamma^3) \leftrightarrow (\gamma^2, \gamma^6), \, \gamma_{4} \mapsto -\gamma_{5},\, \gamma_{5} \mapsto \gamma_{4}, \, \gamma_{7} \mapsto \gamma_{7}. 
\end{align}\end{subequations}
The associated unitary operators read as
\begin{subequations}\begin{align}
   \text{sublattice A:} \quad \sigma_{v} &= \mathcal{U}_{\pi/2}\left(4,2\right) \mathcal{U}_{\pi}\left(4,7\right) \mathcal{U}_{\pi/2}\left(5,1\right) \mathcal{U}_{\pi}\left(5,7\right) \\
   \text{sublattice B:} \quad \sigma_{v} &= \gamma^4\,\mathcal{U}_{\pi/2}\left(1,2\right) \mathcal{U}_{\pi}\left(2,7\right) \mathcal{U}_{\pi/2}\left(4,5\right) \mathcal{U}_{\pi}\left(5,7\right) \mathcal{U}_{\pi/2}\left(3,6\right)\mathcal{U}_{\pi}\left(6,7\right), 
\end{align}\end{subequations}
respectively. Note that $C_{4z}$ follows from $\sigma_d$ and $\sigma_v$.

\subsection{Relation to spin-$S$ operators}

As explained in the main text, the Clifford generators can also be represented in terms of spin operators. For the first model, $\Gamma^{a}$ can be written in terms of spin-$3/2$ operators $S^{j}$ as \cite{Demler1999Jan, Murakami}
\begin{equation}
\label{eq:spin-3/2maptospin}
\Gamma^1=\frac{1}{\sqrt{3}}\left\{S^y, S^z\right\}, \, \Gamma^2=\frac{1}{\sqrt{3}}\left\{S^z, S^x\right\},\,
\Gamma^3=\frac{1}{\sqrt{3}}\left\{S^x, S^y\right\}, \,
\Gamma^4=\frac{1}{\sqrt{3}}\left[\left(S^x\right)^2-\left(S^y\right)^2\right] \,\;\;\text{and} \,\;\;\Gamma^5=\left(S^z\right)^2-\frac{5}{4}\mathbb{I}_4.
\end{equation}

For the second model, one can employ a similar construction using spin-7/2 operators $L^{j}$. In an irreducible spin-$j$ representation, $L_z$ has eigenvalues $m=-j,-j+1,\dots, j.$ The operators
\begin{equation}
P_m\left(L^z\right)=\prod_{ m^{\prime} \neq m} \frac{L^z-m^{\prime}}{m-m^{\prime}},
\end{equation}
are explicit degree-(2$j$) polynomials in $L^z$ and form a basis for all diagonal matrices in the $\ket{m}$ basis. Additionally, any off-diagonal operator can be obtained by applying powers of $L^{+}$ or $L^{-}$, e.g. $\ket{m}\bra{n}\propto P_m(L_z)(L^{+})^{m-n}$. Since $\ket{m}\bra{n}$ form a basis for all matrices, any $8\times8$ matrix, including the Clifford generators $\gamma^{a}$, can be represented as finite-degree polynomials in the spin $7/2$ operators $L^{x},L^{y}, L^{z}$.

\section{Extended phase diagram of the spin-7/2 model and vison spectrum}\label{app:furtherdetails}
\begin{figure}[htp!]
    \centering
    \includegraphics[width=1\linewidth]{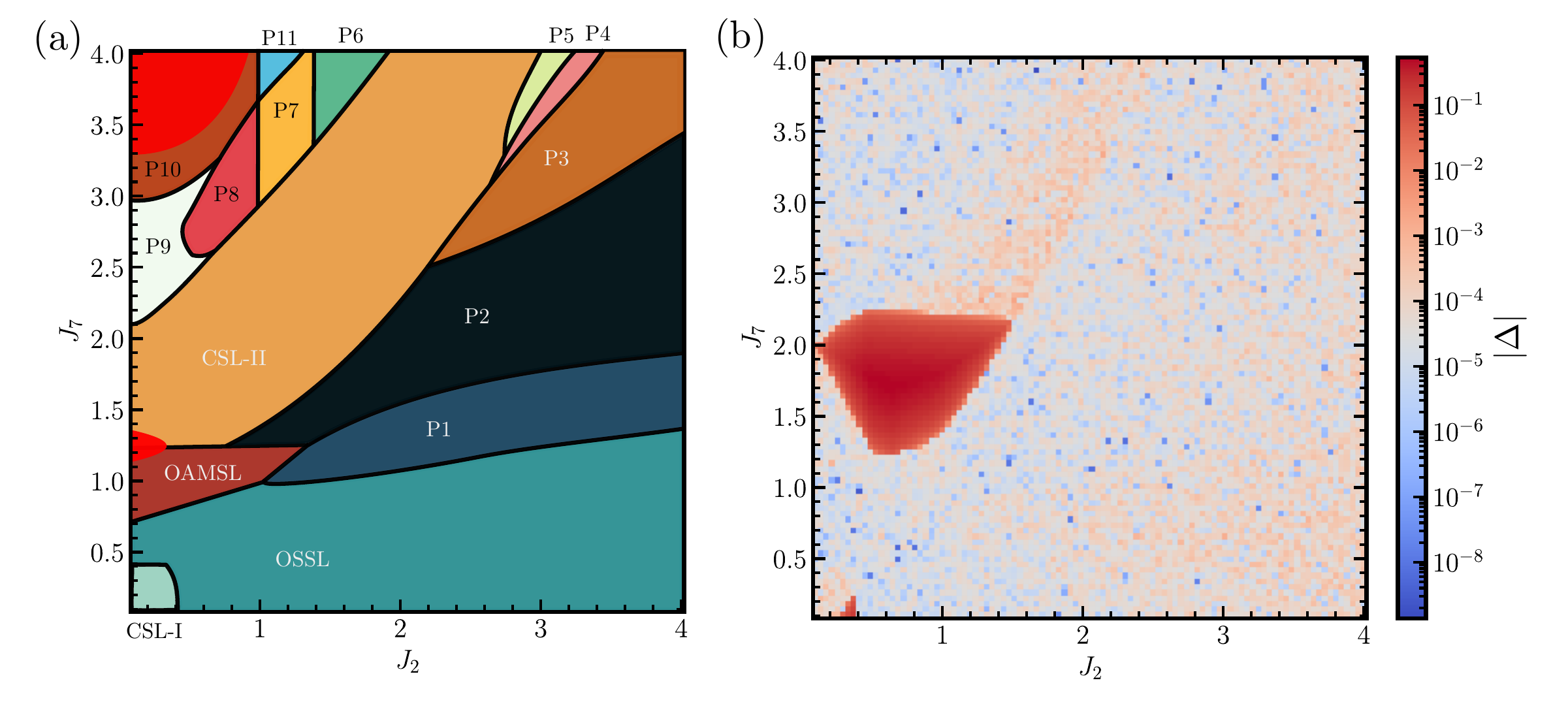}
            \caption{(a) Extended phase diagram for the spin-$7/2$ model shown in \figref{fig:secondmodel}b. Schematic representations of the flux configurations for each phase $P_l$ ($l = 1, \dots, 15$) are shown in \figref{fig:phasesext}. (b) Gap $\Delta$ (log scale) as a function of the couplings for the phase diagram in panel (a). As indicated by the color bar, apart from the CSL-I and CSL-II phases, all remaining phases remain gapless over this parameter range.}
    \label{fig:extendedphasediagram72}
\end{figure}

In \figref{fig:extendedphasediagram72}a, we present the extended phase diagram of the spin-$7/2$ model. As discussed in the main text, increasing $J_{\theta}$ with $\theta = 2, 2^{\prime}, 7$ stabilizes a number of additional phases. These phases are obtained using a super unit cell containing eight sites (indicated by the white square in \figref{fig:secondmodel}c), and their associated phase flux configurations are shown in \figref{fig:phasesext}. 
At $J_2 = J_2^{\prime} = 0$ and for $J_{7} > 2$, the ground state develops a uniform flux configuration, now characterized by a $0$ flux through the square $\emptysquare[45]$ plaquettes. Phases P9 and P10 can be understood as natural extensions of this state, in which additional orbital textures are stabilized by finite values of $J_{2}$ and $J_{2}^{\prime}$.

Rather than introducing new nomenclature for these additional phases, we summarize their flux configurations and highlight their similarities to the phases shown in \figref{fig:secondmodel}b as follows:

\begin{figure}[tb]
    \centering
    \includegraphics[width=1\linewidth]{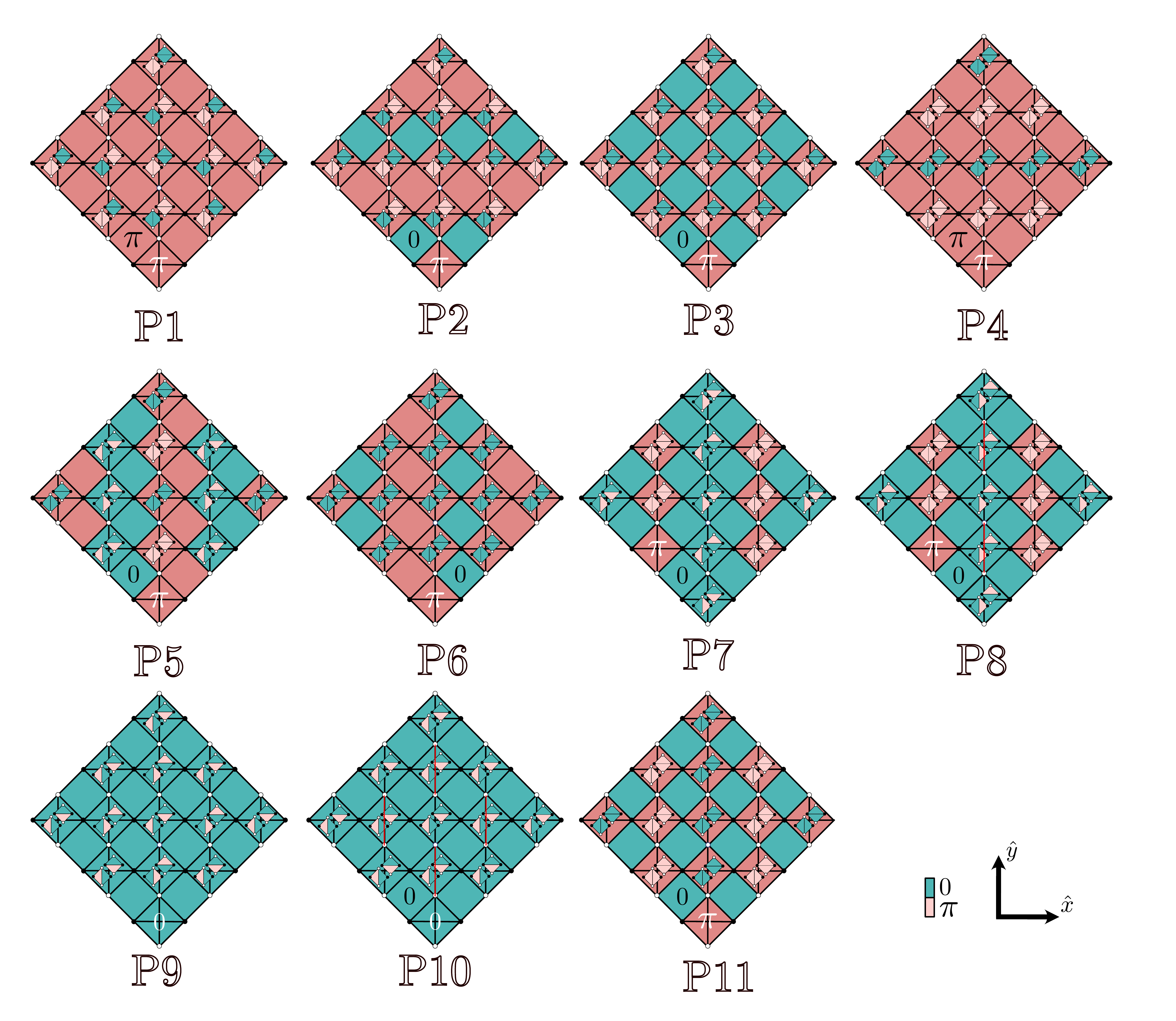}
    \caption{Schematics of flux configurations found in the extended phase  in \figref{fig:extendedphasediagram72}a. 
Elementary plaquettes (empty and crossed squares, and triangles inside the latter) are colored according to their gauge flux values: $0$ (green) or $\pi$ (red). See the discussion in \appref{app:furtherdetails} for more details about each phase. The red bonds in phase P10 (P8) represent the necessary excitations $A$ (see \figref{fig:visons}a) to connect this phase to P9 (P7). }
    \label{fig:phasesext}
\end{figure}

\begin{itemize}

\item P1: Similar to the OSSL phase, with orbital antiferromagnetic stripes now present along both the $\hat{x}$ and $\hat{y}$ directions. The flux configuration is invariant under the combined operation $\Theta \times \mathcal{T}_{x,y}$, where $\mathcal{T}_{x,y}$ denotes translation by one unit cell along the $x$ and $y$ directions.

\item P2: Orbital (A)FM stripes running along the $\hat{y}$ direction, with staggered $\pi$ and $0$ flux stripe patterns along ($\hat{y}$)$\hat{x}$ on the $\emptysquare[45]$ plaquettes, and $\pi$ flux on the $\squarecross[45]$ plaquettes. 

\item P3: Orbital textures arranged as in the OAMSL (d-wave structure), with a checkerboard pattern of $0$ and $\pi$ fluxes on the $\emptysquare[45]$ and $\squarecross[45]$ plaquettes, similar to the CSL-II flux configuration. The flux configuration is invariant under the combined operation $\Theta \times C_{4z}$.

\item P4: Orbital stripes staggered along the $\hat{y}$ direction with a uniform $\pi$ flux over the square $\squarecross[45]$ and $\emptysquare[45]$ plaquettes. The flux configuration is invariant under the combined symmetry $\Theta \times \mathcal{T}_y$.

\item P5: The flux configuration breaks $\Theta, \mathcal{T},\sigma_d$ and does not exhibit a simple stripe or checkerboard structure, indicating a more complex orbital ordering, which could be related to this phase appearing at the border of phases CSL-II, P4 and P3 in the extended phase diagram. 

\item P6: Uniform orbital textures ($\pi$  flux), with staggered $\pi$ and $0$ flux stripes patterns along $\hat{y}$ on the $\emptysquare[45]$, and  $\pi$ flux on the $\squarecross[45]$ plaquettes.

\item P7: A complex phase with alternating orbital textures $(+,-,+,-)\rightarrow (-,+,-,+)$ along the $\hat{y}$ direction and uniform stripes $(-,-,-,-)$ along the $\hat{x}$ direction. The $\emptysquare[45]$ plaquettes host $0$ flux, while the $\squarecross[45]$ have a stripes of alternating $0$ and $\pi$ fluxes along the $\hat{x}$ direction.

\item P8: Same square and orbital flux pattern as in P7 apart from the fluxes influenced by a flip of the indicated red bonds.  

\item P9: Uniform $0$ flux on all square plaquettes, accompanied by an alternating pattern of $(+,-,+,-)\rightarrow (-,+,-,+)$ on the orbital textures along both directions. The flux configuration is invariant under the combined operation $\Theta \times \mathcal{T}_{x,y}$.

\item P10: Same square and orbital flux pattern as in P9 apart from the fluxes influenced by a flip of the indicated red bonds, i.e., instead of the alternating pattern, orbital textures have a fixed $(+,-,+,-)$ structure. The flux configuration is invariant under the combined operation $\Theta \times \sigma_d$. Furthermore, the orbital flux structure has a p-wave structure.

\item P11: Same square flux patterns as in P3 and the CSL-II. Additionally, hosts staggered orbital stripes of uniform fluxes along the $\hat{y}$ direction. The flux configuration is invariant under the combined operation $\Theta \times \mathcal{T}_x$.
\end{itemize}

\begin{figure}[htp!]
    \centering
    \includegraphics[width=1\linewidth]{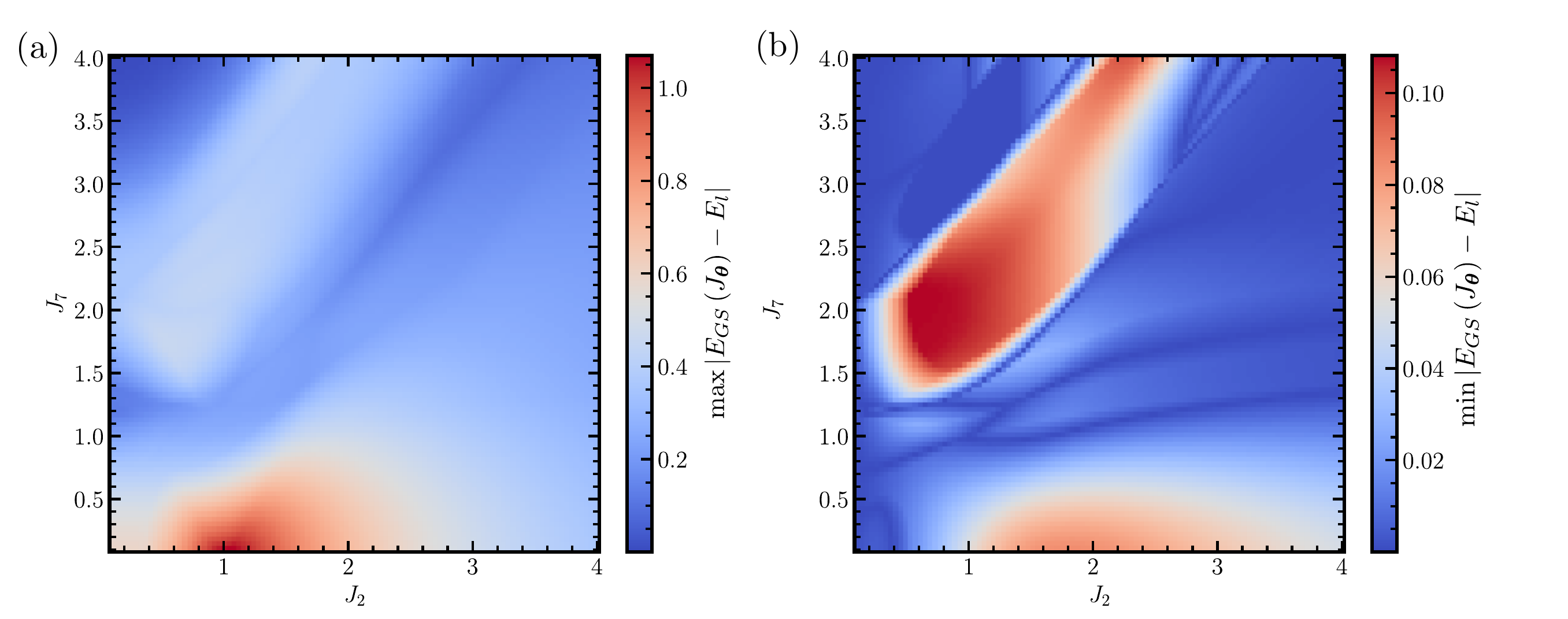}
            \caption{Maximum (a) and minimum (b) energy differences between the ground state at each $J_{\{2,7\}}$, for $J_{2}^{\prime} = J_{2}$ and $J_{1}^{\prime} = J_{1}=1$, and the remaining energies $E_l$ corresponding to each phase in \figref{fig:extendedphasediagram72}a.}
    \label{fig:extendedphasediagram72-2}
\end{figure}

By computing the difference between the ground-state energy $E_{\mathrm{GS}}$ at each $J_{\theta}$ and the energies $E_l$ of the remaining phases ($l = 1, \dots, 15$), we can identify the energy boundaries between phases.
In particular, the maximum of this energy difference (\figref{fig:extendedphasediagram72-2}a) reveals regions where the phases are energetically close, with two such regions highlighted in red in \figref{fig:extendedphasediagram72}a, indicating potential degeneracies. Taking instead the minimum of this quantity (\figref{fig:extendedphasediagram72-2}b) allows us to identify the energetically nearest competing phase.

For example, phases P9 and P10, as well as P8 and P7, are very close in energy, which can be understood from their gauge-flux configurations. As shown in \figref{fig:phasesext}, the red bonds in phases P10 (P8) indicate the bond flips required to reach phases P9 (P7).

\begin{figure}[tb]
    \centering
    \includegraphics[width=1\linewidth]{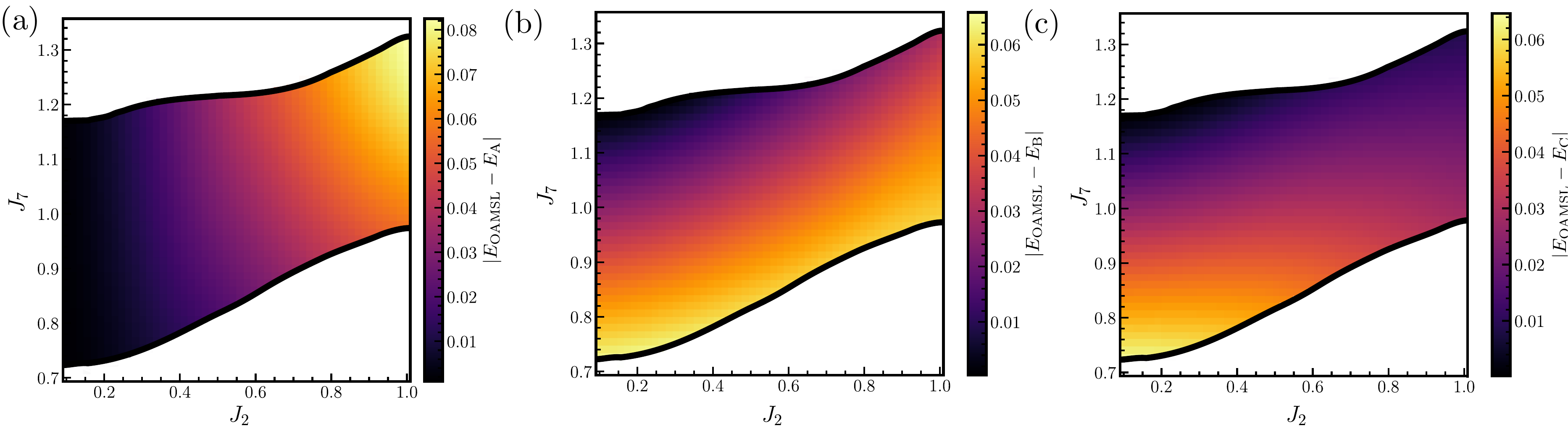}
    \caption{Energy distribution of the possible excitations for the OAMSL in the spin-7/2 model (see \figref{fig:visons}): the non-topological excitation A (a), visons B (b) and C (c) as a function of $J_{2}$ and $J_{7}$ for 
            $J_{2}^{\prime}=J_{2}$ and $J_{1}^{\prime}=J_{1}=1$. The colorbar indicates the energy difference from each excitation (X) to the OAMSL in $\left|E_{\text{OAMSL}}-E_{\text{X}}\right|$.}
    \label{fig:supvison}
\end{figure}

All phases appear to be gapless—within the $100 \times 100$ momentum-grid resolution used in our simulations—except for the CSL-I and CSL-II phases, which become clearly gapped in certain regions of the phase diagram (\figref{fig:extendedphasediagram72}c). The degeneracy at the boundary between the CSL-II and the OAMSL phase is also reflected in the vison spectrum (see \figref{fig:supvison}), where all excitations exhibit a small gap relative to the ground-state energy.

\end{appendix}

\end{document}